\documentclass[draft]{agujournal2019}
\usepackage[]{gensymb,amsmath,amssymb}
\usepackage{caption}
\usepackage{subcaption}
\usepackage{url} 
\usepackage{lineno}
\usepackage[inline]{trackchanges} 
\usepackage{soul}

\newcommand{\ts}[1]{\textsuperscript{#1}}

\newcommand{\sUnit}{s}

\newcommand{\productionCharge}{pc}
\newcommand{\productionDischarge}{pd}
\newcommand{\totalEnergy}{pe}
\newcommand{\effCharge}{\eta^{c}}
\newcommand{\effDischarge}{\eta^{d}}
\newcommand{\maxCharge}{\overline{PC}}
\newcommand{\maxDischarge}{\overline{PD}}
\newcommand{\maxEnergy}{\overline{PE}}
\newcommand{\minEnergy}{\underline{PE}}

\newcommand{\unit}{g}
\newcommand{\unitSet}{G}
\newcommand{\node}{n}
\newcommand{\nodeSet}{N}
\newcommand{\nodeInjection}{inj}
\newcommand{\timeUnit}{t}
\newcommand{\timeSet}{T}
\newcommand{\flow}{f}

\newcommand{\production}{p}

\newcommand{\productionMax}{\overline{P}}
\newcommand{\productionMin}{\underline{P}}
\newcommand{\rampUp}{RU}
\newcommand{\rampDown}{RD}

\newcommand{\demand}{D}

\newcommand{\unitCommit}{u}
\newcommand{\decisionStart}{v}

\newcommand{\decisionStop}{w}

\newcommand{\minUpTime}{UT}
\newcommand{\minDownTime}{DT}

\newcommand*\rot{\rotatebox{90}}
\draftfalse

\journalname{Earth's Future}

\begin{document}

%
%


\title{Linking Unserved Energy to Weather Regimes}

%
%




\authors{Rogier H. Wuijts\affil{1,2}\thanks{These authors contributed equally to this work.}, 
Laurens P. Stoop\affil{2,1,3}\thanks{These authors contributed equally to this work.}\thanks{0000-0003-2756-5653}, 
Jing Hu\affil{1,4}\thanks{0000-0002-1182-5687}, 
Arno Haverkamp\affil{3}\thanks{0000-0001-6947-7892}, 
Frank Wiersma\affil{3}, William Zappa\affil{3}\thanks{0000-0001-6810-7224}, 
Gerard van der Schrier\affil{5}\thanks{0000-0001-7395-8023}, 
Marjan van den Akker\affil{2}\thanks{0000-0002-7114-0655}, 
Machteld van den Broek\affil{6}\thanks{0000-0003-1028-1742}}

\affiliation{1}{Copernicus Institute of Sustainable Development, Utrecht University, the Netherlands}
\affiliation{2}{Information and Computing Science, Utrecht University, the Netherlands}
\affiliation{3}{TenneT TSO B.V., Arnhem, the Netherlands}
\affiliation{4}{PBL Netherlands Environmental Assessment Agency, the Netherlands}
\affiliation{5}{Royal Netherlands Meteorological Institute, the Netherlands}
\affiliation{6}{Integrated Research on Energy, Environment and Society, Energy and Sustainability Research Institute Groningen (ESRIG), University of Groningen, the Netherlands}




\correspondingauthor{Laurens Stoop}{l.p.stoop@uu.nl}





\begin{keypoints}
\item Our research explores the relationship between weather regimes and Energy Not Served (ENS) events in Europe.
\item ENS events in central European countries often coincide with two weather regimes associated with cold and calm weather conditions.
\item The weather regime during the preceding 10 days is an indicator for a ENS event, showing the dynamic component of the energy system. 
\end{keypoints}

%
%

%
%


\begin{abstract}
The integration of renewable energy sources into power systems is expected to increase significantly in the coming decades. This can result in critical situations related to the strong variability in space and time of weather patterns. During these critical situations the power system experiences a structural shortage of energy across multiple time steps and regions, leading to Energy Not Served (ENS) events.

Our research explores the relationship between six weather regimes that describe the large scale atmospheric flow and ENS events in Europe by simulating future power systems. Our results indicate that most regions have a specific weather regime that leads to the highest number of ENS events. However, ENS events can still occur during any weather regime, but with a lower probability.

In particular, our findings show that ENS events in western and central European countries often coincide with either the positive Scandinavian Blocking (SB+), characterised by cold air penetrating Europe under calm weather conditions from north-eastern regions, or North Atlantic Oscillation (NAO+) weather regime, characterised by westerly flow and cold air in the southern half of Europe. Additionally, we found that the relative impact of one of these regimes reaches a peak 10 days before ENS events in these countries. In Scandinavian and Baltic countries, on the other hand, our results indicate that the relative prevalence of the negative Atlantic Ridge (AR-) weather regime is higher during and leading up to the ENS event.

\end{abstract}

\section*{Plain Language Summary}
With the energy transition, energy system evolve and renewable energy technologies such and solar PV and wind turbines become more common. This increases the relation between weather, climate and the energy systems. 

To assess the reliability of a possible evolution of an energy system, we can look at possible structural shortage of energy supply due to either the weather or the system design. For this purpose, an Energy System Model can be developed in which the properties of a specific energy system design are accounted for. Such Energy System Models provide insight into a possible operation of the energy system and thereby insight into its reliability. 

In our research we found that for Europe possible structural shortage due to the weather occurs in the winter. In the winter the complex state of the weather can be simplified into six states. We found that most regions have a specific state in which there is an increased risk of possible structural shortage. This riskiest state for central Europe is associated with colder temperatures and calm weather. The riskiest state for northern Europe is associated with sunny, but cold weather in north-eastern Europe.

%
%

\section{Introduction}

In 2022, anthropogenic greenhouse gas emissions are estimated to already have caused approximately 1.0°C of warming to the average global surface temperature compared to pre-industrial levels (1850–1900)~\cite{RN2}. To avoid the worst consequences of climate change, the world strives to rapidly reduce its greenhouse gas emissions~\cite{parisAgreement,greenDeal}. In order to reduce CO$_2$ emissions in the power system, supply needs to shift to renewable energy sources (RES) complemented by low carbon generators~\cite{WEO}. In the future a large part of RES are variable such as solar photovoltaic (PV), onshore and offshore wind, and run-of-river hydro power. The relative share of RES in the global production of electricity is increasing, 10.5\% of the total generation of electricity in 2021 came from solar and wind compared to less than 1\% in 2012~\cite{BNEF}. Moreover, more than 80\% of the newly installed capacity in power systems around the world in 2021 came from RES~\cite{irena2022world}.
However, the energy production of these sources is uncertain and variable. To mitigate this variability a power system must have sufficient storage, demand side response, low/non carbon emitting supply, and transmission capacity to spread the uneven power generation over time and space. 

A critical situation in a power system may not always manifest as a high residual load at a single time step and place~\cite{vanderwiel2019extreme,vanderwiel2019regimes,craig2022disconnect}, but can also manifest as a structural shortage of energy over multiple time steps and across multiple regions~\cite{stoop2021detection}. When these critical situations occur the power system cannot supply every demand, resulting in Energy Not Served (ENS).

ENS is an important reliability indicator for power systems. ENS is the part of demand which is not supplied in a given region over a given time period due to insufficient supply or demand-side resources, implying the Transmission System Operator (TSO) would need to curtail demand involuntary to maintain stable system operation. The expected Energy Not Served (EENS) is one of two key reliability indicators which must be calculated in the European Union as part of the European Resource Adequacy Assessment (ERAA)~\cite{ERAA2021}. The other key reliability indicator is loss of load expectation (LOLE), i.e., the average number of hours ENS that is expected to occur per year based on simulations. However, in power system models, LOLE can take on arbitrary values making it a less robust indicator than EENS~\cite{wuijts2022pitfalls}.

Weather regimes are classifications of the European winter time period meteorological variability
\footnote{Meteorological or climate variability describes variations and changes in the mean state \emph{and} other aspects of climate. Climate variability occurs due to natural and sometimes periodic changes in the circulation of the air and ocean, volcanic eruptions, and other factors. This variability ranges over all spatial and temporal scales, from localised thunderstorms, to larger-scale storms or droughts, and from day-to-day to multi-year, multi-decade and even multi-century time scales.} at the synoptic scale\footnote{In meteorology, the synoptic or large scale is used to indicate weather systems ranging in size from several hundred to several thousand kilometres. This corresponds to a horizontal scale typical of mid-latitude high pressure systems, extra-tropical cyclones and storms.} into quasi-stationary, persistent and recurrent large-scale atmospheric circulation patterns. As the weather in the winter period in Europe is more persistent, weather regimes are often defined for the extended winter~\cite{michelangeli1995weather,neal2016flexible,falkena2020regimes}, although other year-round definitions exist~\cite{Grams2017weather}. The circulation pattern over Europe, and thus a specific weather regime, influences the renewable generation and the energy demand in Europe. Therefore, it may be more difficult to supply all energy demand in certain weather regimes than others, potentially leading to ENS~\cite{vanderwiel2019regimes,Bloomfield2019regimes,Otero2022,tedesco2023}. 

In this article, we aim to investigate the relationship between ENS and weather regimes in Europe. Specifically, we want to determine which weather regimes lead to the highest levels of ENS in European countries. By identifying these critical weather regimes, we can better understand the factors that contribute to ENS and find directions to make the future European power system more resilient. We explore this relationship by simulating the scenarios of a future European power system created by ENTSO-E, incorporating 28 historical weather years in total. For these simulated years we calculate when and how much ENS occurs for each time step and region and if this ENS concurrently occurs with specific weather regimes.

\section{Methods \& Modelling}
In this section, we provide an overview of our methods. We first describe the baseline capacity scenarios we used to model the future European power system. Next, we explain how we constructed our power system model and how we calculated weather-dependent variables, such as wind and solar generation and hydro inflow, for multiple weather years. In the following section we outline the weather regime classification we employed and finally we explain how we generated weather-dependent demand.

\subsection{Energy System Scenarios}
As basis of our power system model, we used the future European power system scenarios from the 2020 Ten Year Network Development Plan (TYNDP2020) created by the European Network of Transmission System Operators for Electricity (ENTSO-E) and Gas (ENTSO-G)~\cite{entso2020tyndp2020}. These scenarios provide insights into the possible energy system of the future and the effects of changes in supply and demand in the energy system.

For our study we used the three scenarios of the TYNDP2020 study as a starting point, namely the three different pathways of National Trends (NT), Global Ambition (GA) and Distributed Energy (DE) for the target year 2040. 
Table~\ref{tab:TYNDPScenario} gives an overview of the total generation capacity per technology for each scenario. 
Each scenario consists of 55 ‘nodes’ corresponding roughly to the current bidding zones in Europe. Each bidding zone usually covers an entire country except for Norway, Denmark, Sweden, and Italy as these are divided into multiple zones. In \ref{app:regioncodes} an overview of the bidding zones, their region code, and the corresponding countries is provided. \ref{app:specificCAP} presents an overview of the installed capacities for some bidding zones of the DE scenario as this is mainly used in this analysis.

\begin{table}[ht]
\caption{Total capacity (GW) in Europe for different generation technologies for the different TYNDP Scenarios in 2040. }
\label{tab:TYNDPScenario}
\begin{tabular}{c|ccccccccc}
       & \rot{Thermal (Coal, Nuclear etc.)} & \rot{Run of River} & \rot{Closed pumped hydro storage} & \rot{Open pumped hydro storage} & \rot{Hydro Storage} & \rot{Solar Photovoltaic} & \rot{Offshore Wind} & \rot{Onshore Wind} & \rot{Battery} \\ \hline
Distributed Energy (DE) & 431           & 67        & 42               & 62             & 109       & 750            & 79                  & 585                & 112           \\
National Trends (NT) & 451           & 67        & 42               & 62             & 109       & 489            & 131                 & 406                & 61            \\
Global Ambition (GA) & 441           & 67        & 42               & 62             & 109       & 437            & 146                 & 440                & 40           
\end{tabular}
\end{table}

The TYNDP scenarios are comprehensive datasets that provide a detailed breakdown of the assumed generator capacities of different technologies in the different bidding zones. The datasets include generators of various fuel types and ages, which can impact their efficiency. However, technological details such as average unit size and ramping limits have been added from other sources~\cite{PONCELET2020113843,etri2014energy}, as these were not included in the TYNDP datasets. The transmission capacity between regions is given by net transfer capacity (NTC) values, which are constant throughout the year but differ per scenario. All the capacity and grid data used and information about their origin is included in an online dataset\footnote{The capacity data and its source can be found here \url{https://github.com/rogierhans/TYNDP2040ScenarioData}.}.

\begin{figure}
    \centering
    \includegraphics[scale=0.65]{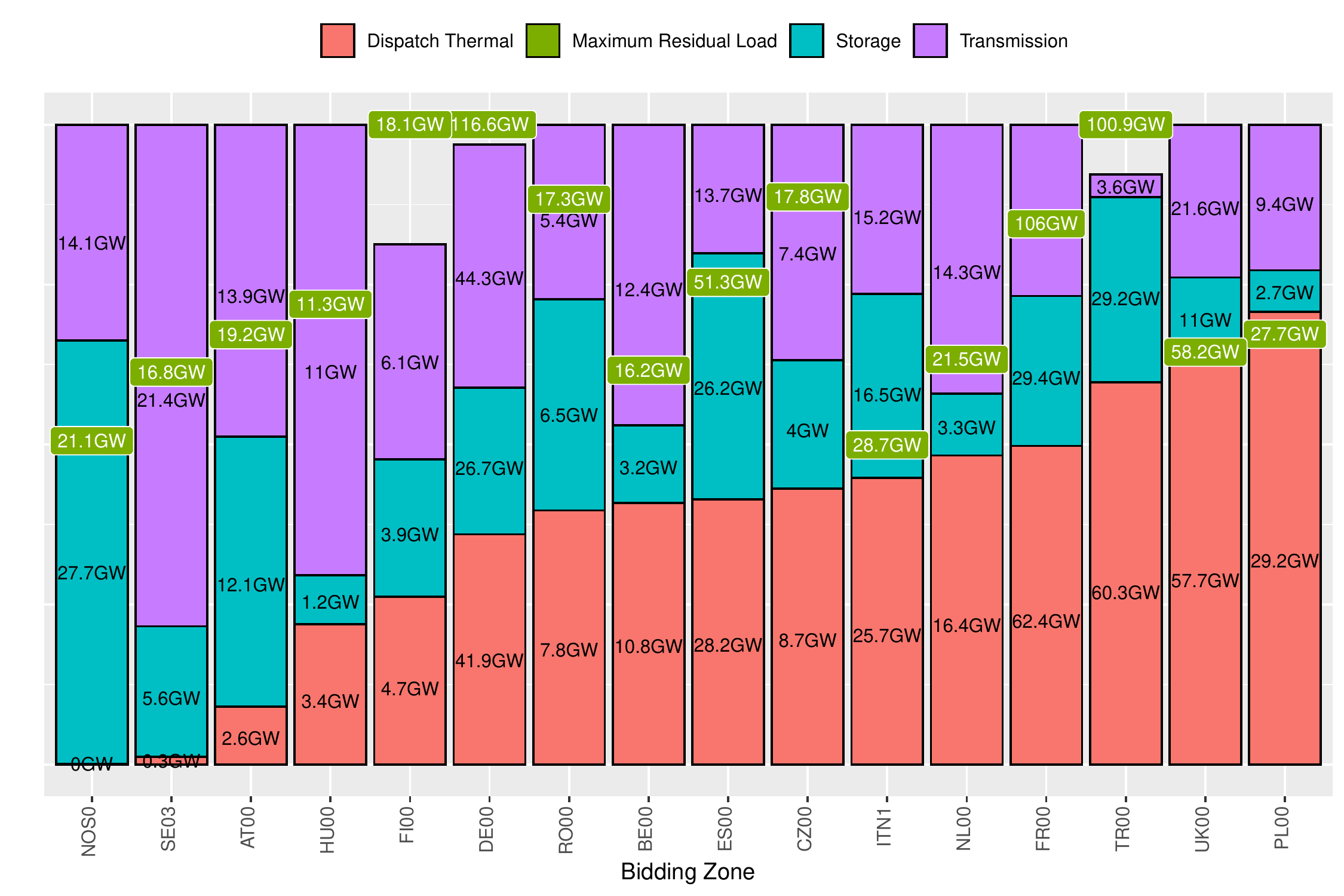}
    \caption{As a percentage of total capacity in the region, the total thermal dispatch capacity (red), the additional transmission capacity (purple), and storage capacity (blue) are shown for a subset of bidding zones with high residual load. In addition, the maximum residual load (green box) over 28 weather years for the scenario \textit{Distributed Energy 2040} is shown. See \ref{app:regioncodes} for the explanation of the codes. }
    \label{fig:MaxResload}
\end{figure}

The maximum residual load is defined as the remaining demand when renewable energy generation (i.e. from solar, wind and run-of-river hydro) is fully utilised to meet the demand. We found that for some bidding zones and scenarios, the maximum residual load exceeds the total thermal, transmission and storage capacity for that zone (Figure \ref{fig:MaxResload}). However, most bidding zones in these scenarios are adequate by design \cite{entso2020tyndp2020}. In addition, we do not utilise the demand side response options that were defined to negate this. 

To identify critical situations in which not all demand can be met by generation, we made the TYNDP scenarios more challenging by either increasing the demand (i.e. an increase in electrification) or decreasing the generation (i.e. a decrease in carbon-emitting generators). We define 12 scenarios: four alternatives of each of the 3 TYNDP scenarios: (i) unchanged, (ii) +10\% demand, (iii) +20\% demand and (iv) -20\% generation capacity.  For the latter, we only decrease the capacity of highly emitting generators that use as fuel either hard coal, lignite, or oil.

\subsection{Power System Model}\label{sec:psm}
We used the unit commitment and economic dispatch problem (UCED) to simulate the hourly operation of the power system in Europe. The UCED is a mathematical optimisation model that finds the most cost-effective schedule for operating generators, ensuring that all demand is met at every time step and in every region. This method was chosen for its ability to simulate hourly market operations with high technical detail, which is commonly used in the power industry~\cite{welsch2014incorporating,abujarad2017recent}.

The UCED may incorporate many types of technical constraints, of which some can be omitted depending on the purpose of the study~\cite{wuijts2022modelchar}. We therefore can use a simplified model which includes fewer constraints regarding the flexibility of thermal power plants. Moreover, we explicitly minimise the ENS in the power system and not the total system cost, which implicitly minimises the same amount of ENS. The complete model specification, our simplifications and the validation of the assumptions underlying these simplifications can be found in \ref{appendix:PSM} and \ref{app:validationPSM} respectively.

We simulate the hourly operation of the power system for a total of 28 weather years (1982-2010) in yearly segments from the 1\ts{st} January to the 31\ts{st} of December, i.e. 8760 hours in one model run. We set the storage levels at the start of the simulation at half capacity and enforce that the capacity is also at half capacity at the end of the year. This ensures that within a simulated year any storage discharged, e.g. hydro reservoirs are compensated by pumping or hydro inflow during that year.

\subsection{Modelling weather dependent variables}
The TYNDP scenarios contain, among others, technologies that depend on the weather, see Table~\ref{tab:TYNDPScenario}. When the influence of the weather on an energy system is investigated, it is vital that the potential generation of these technologies is modelled accurately~\cite{craig2022disconnect}. As the spatial distribution of measurements is sparse, climate model data is utilised. As the data for the different technologies was collected from various sources, care was taken to make sure that the underlying climate model is consistent across them. 

For all weather dependent variables, data derived from the ERA5 reanalysis dataset is used~\cite{hersbach2018era5}. Wind and solar energy generation was determined with the conversion models as described in Section \ref{section:RESgeneration} for the period 1950-2022. The hydrological data provided through by the E-Hype project, as described in Section \ref{section:HydroInflow}, was only available from 1980-2010. The energy demand dataset, as described in Section \ref{section:Demand}, was only available from 1982-2016. The analysis is, therefore, limited to the period 1982-2010, as this is the period with maximum overlap.

\subsubsection{Energy conversion models for Wind and solar}\label{section:RESgeneration} 
To determine the electricity generation time series of RES per bidding zone, two things need to be determined. First, the potential generation profile per unit of installed capacity for solar photovoltaic and wind turbine technology must be determined for each climate model grid cell. Second, the potential generation of each technology must be multiplied by the distribution of installed capacity in each grid cell and summed up over all grid cells within a bidding zone. 

Conversion models can be employed to calculate the potential generation of wind and solar energy. In previous work within the ACDC-ESM project~\cite{stoop2021detection}, a number of conversion models were analysed and compared together with the TSO stakeholder, TenneT TSO B.V., to determine which were representative yet computationally simple. For solar photovoltaic electricity generation we follow these recommendations and use the relatively simplified method as described by~\citeA{Jerez2015}. For comprehensiveness, \ref{appendix:Solar} provides the exact description of the method used. More elaborate methods, like the one presented by \citeA{SaintDrenan2018}, were not used, as these require additional information on panel tilt angle and solar radiation components that are not available within the TYNDP2020 scenario building guidelines.

For wind turbine electricity generation, we follow the recommendation laid out by \citeA{stoop2021detection} and use the method as described from \citeA{Jerez2015}. However, we made four adjustments to this model to align it with the TYNDP2020 study. First, we reduced the effective maximum capacity factor ($CF_e$) by $5\%$ to $95\%$ to represent the wake and array losses in large scale wind-farms~\cite{Lundquist2018,Bleeg2018,Fischereit2021,Saint-Drenan2020}. Secondly, we let the capacity factor linearly decrease at high wind speeds to more accurately represent high windspeed operational conditions. The third change was that we tuned the power curve regimes based on stakeholder input. Finally, the wind speed provided by ERA5 (at 100 meter) does not match the hub heights of turbines within the TYNDP scenarios used, therefore it is scaled using the wind profile power law to 150 meters for offshore turbines and 120 meters for onshore turbines. See \ref{appendix:Wind} for the formal definitions and more detailed discussion of the methods used. 

For both technologies the total energy generation per bidding zone per hour was obtained by multiplying the weighted mean generation profile per bidding zone with the installed capacity as provided by the \citeA{entso2020tyndp2020} for the scenario used. The weights in the averaging procedure for each bidding zone were determined by the mean capacity factor of each grid cell within that zone. Grid cells that were partially within a bidding zone got an additional weighting based on the percentage of the area within a zone. 

\subsubsection{Hydro inflow data}\label{section:HydroInflow}
The  hydro inflow data is based on historical river runoff reanalysis data simulated by the E-HYPE model~\cite{donnelly2016using}. E-HYPE is a pan-European model developed by The Swedish Meteorological and Hydrological Institute (SMHI), which describes hydrological processes including flow paths at the subbasin level. E-hype only provides the time series of daily river runoff entering the inlet of each European subbasin over 1980-2010. To match the operational resolution of the dispatch model, we linearly downscale these time series to hourly. By summing up runoff associated with the inlet subbasins of each country, we also obtain the country-level river runoff. 

The hydro inflow time series per country as inputs of the UCED model is defined as the normalized energy inflows (per unit installed capacity of hydropower) embodied in the country-level river runoff. The dispatch model decides whether the energy inflows are actually used for electricity generation, stored, or spilled (in case the storage reservoir is already full). Specific details on the modelling method can be found in \ref{app:hydropowermodel}.

We explicitly consider three types of hydropower plants, namely storage hydropower plant (STO), run-of-river hydropower plant (ROR) and pumped storage hydropower plant (PHS). For modelling purposes, we need to estimate the specific maximum energy storage content, for each type of hydropower. We obtain this by using an in-house database, containing the information of 207 hydropower plants, and calibrating this with present level of total storage size (220 TWh) in Europe given by \citeA{mennel2015hydropower}.

\subsection{Weather Regimes}\label{sec:weatherregimes}
In this study we use the classification of the atmospheric state from \citeA{falkena2020regimes}. They revisited the identification of European weather regimes and showed that six clusters should be used in the classification, see \ref{app:weatherregimes}. The six weather regimes used are defined for the whole of Europe at a daily interval for the period December-March from 1979-2018, but we only used the period 1982-2010 due to the availability of other datasets. The six regimes used have been labelled to indicate atmospheric state. Due to their symmetry, a name (Atlantic Ridge (AR), North Atlantic Oscillation (NAO) and Scandinavian Blocking (SB)) and state ($+$ and $-$) are used to label each of the six weather regimes. It should be noted that the naming convention used, does not imply that the weather associated with these six weather regimes is similar to a definition that uses four or two clusters to classify the weather. 

While the specific weather can vary within a weather regime, the general flow of air is consistent within a weather regime, see also \ref{app:metWR}. The AR+ regime is associated with north-to-south air flow, sunny weather in north-eastern Europe and with calm weather in the Atlantic. The AR- regime is associated with sunny, but cold weather in north-eastern Europe, with a gradient to slightly warmer weather in the south-west. Under both AR regimes south-west Europe is characterised by decreased wind speeds and increased solar irradiance. The NAO+ regime is associated with a westerly flow, bringing warm temperatures and higher wind speeds to Scandinavia, while in the south colder temperatures and less sunlight is seen. Warmer and sunnier weather is observed in most of Europe during the NAO- state, due to south-)easterly flow. On the other hand, the blocking pattern of the SB+ regime, due to a stationary large high pressure system in Scandinavia, generally brings cold and calm weather to central and northern Europe. While the SB- regime shows an increased wind speed in central Europe and the Atlantic. Both SB regimes show strongly reduced solar irradiance. 

The persistence and occurrence of a specific weather regime is subject to decadal variability~\cite{Dorrington2022}, and thus depend on the analysed period. Therefore, we show in Table~\ref{tab:WRTable} the occurrence (the presence of weather regimes in the total number of days), how many times a persistent period of that weather regime occurs, the persistence (the average length) and maximum length for the six weather regime used. 

\begin{table}[]
\centering
\caption{Prevalence of the daily defined European weather regimes (WR) in the winter months (December, January, February, March)
from 1982 to 2010: the number of days the WR occurs, number of periods (consecutive days with the same WR), the average and
maximum length of these periods.}
\label{tab:WRTable}
\begin{tabular}{c|cccc}
WR   & number of days & number of periods & avg length period & max length period \\ \hline
AR-  & 550 (15.6\%)   & 114               & 4.8 days              & 23 days       \\
AR+  & 562 (16.0\%)   & 151               & 3.7 days              & 16 days       \\
NAO- & 611 (17.4\%)   & 145               & 4.2 days              & 20 days       \\
NAO+ & 697 (19.8\%)   & 167               & 4.2 days              & 16 days       \\
SB-  & 568 (16.2\%)   & 157               & 3.6 days              & 22 days       \\
SB+  & 528 (15.0\%)   & 89                & 5.9 days              & 32 days              
\end{tabular}
\end{table}

\subsection{Weather dependent demand}\label{section:Demand} 
Weather not only  influences the generation of electricity, but also the demand for electricity, primarily for heating, cooling and lighting. The effect of temperature based metrics like Heating Degree days on demand are known~\cite{Quayle1980} and well established metrics in impact assessments~\cite{vanderwiel2019extreme,bloomfield2021quantifying}. Scenarios for the future like the TYNDP2020, take into account system wide changes to an energy system. Not only the influence of temperature on the need for space heating and cooling related demand is taken into account, but also the transition to heat pumps and the additional demand from  transport electrification (private cars, busses, passenger trains and heavy goods)~\cite{entso2020tyndpguidelines}. 

In this study we use the hourly weather dependent demand time series for each bidding zone and scenario that were generated for the TYNDP2020. As the weather model that was used to obtain the Pan-European Climate Database version 3.1 (PECDv3.1) dataset is ERA5, the driving weather is consistent with the other data sources used. Because, the PECDv3.1 dataset initially published by ENTSO-E contained a few errors (countries missing data under specific scenarios), the specific updated and complete datafiles used were provided by ENTSO-E through the ACDC-ESM project.

\section{Results}
In this section, we present our results in three stages. We first show the link between energy not served (ENS) and weather regimes for the 12 adjusted TYNDP scenarios. Then we discuss to what degree the sequence of weather regimes increases or reduces the risk for the energy system. And finally, we look deeper into what the meteorological link is between the weather regimes and ENS.

In the analysis we focus on two typical regions based on subsets of bidding zones. The first regions consist of  Germany (\textit{DE00}), France (\textit{FR00}) and the Netherlands (\textit{NL00}) and represents central Europe. The second regions consist of Lithuania (\textit{LT00}), Latvia (\textit{LV00}), the southern region of Norway (\textit{NOM1}) and the northern region of Sweden (\textit{SE01}) and represents the Scandinavian and Baltic region.

\subsection{Scenario dependency}\label{sec:resultScenario}
\begin{figure}[!htpb]
    \centering
    \includegraphics[width=0.8\textwidth]{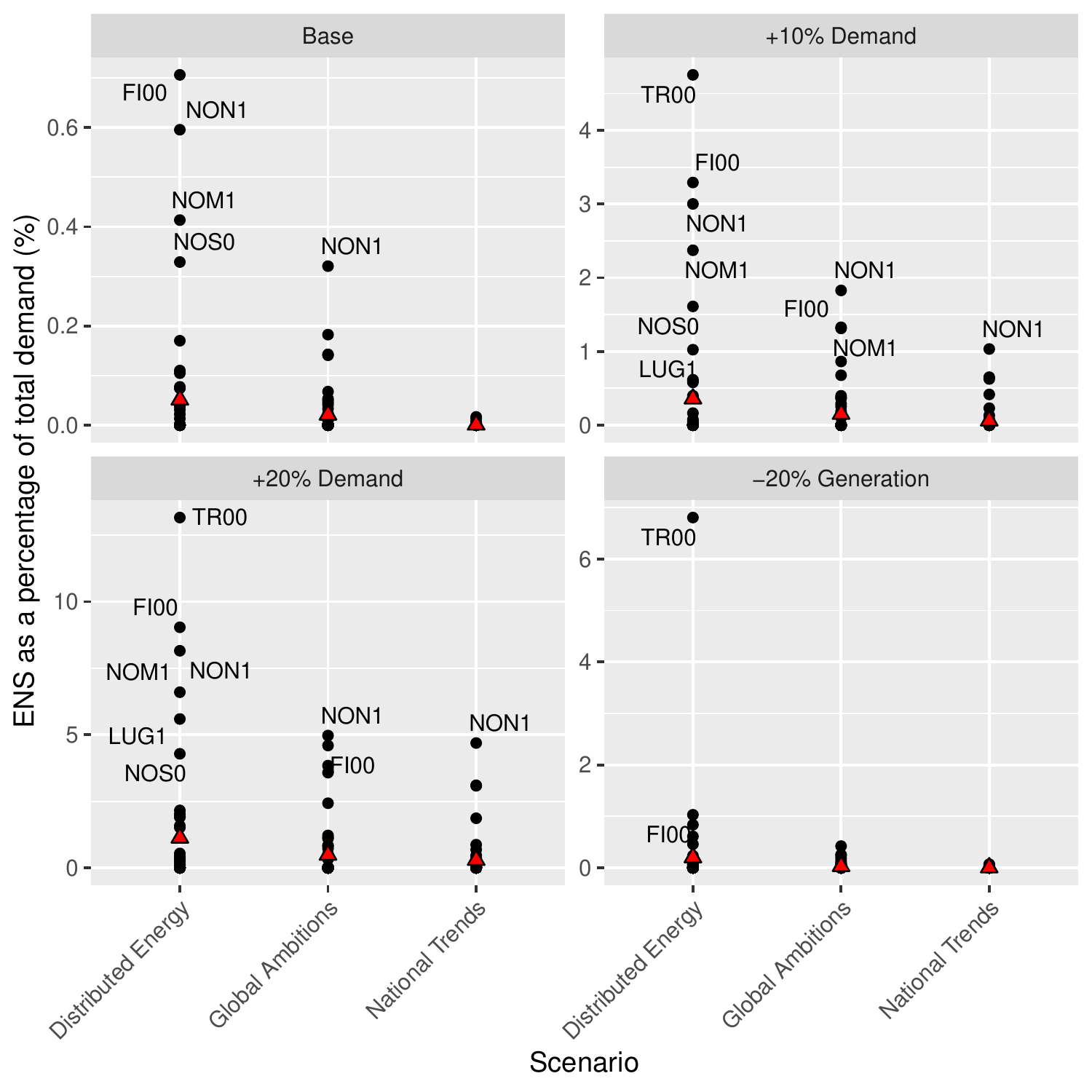}
    \caption{The Energy Not Served as a percentage of the total electricity demand of each bidding zone based on weather years 1982 to 2010. The red triangle represents the average over all bidding zones. The twelve scenarios are clustered by their variation. The regions with a high share of ENS are labelled with their region code. Please note that the \emph{y-axes are on different scales}. }
    \label{fig:ENSPercentage}
\end{figure}

In Figure~\ref{fig:ENSPercentage} the ENS as a percentage of the total demand is shown for all twelve scenarios. We can see that for the base scenarios of the TYNDP, Finland (FI00) and Norway (NON1, NOS0, NOM1) have some ENS (approximately $0.6$\% of the total demand), but for most countries the ENS is close to zero.  When the base scenario is stretched, either by increasing the demand or decreasing the generation capacity of dispatchable generators, then we indeed see more ENS. Especially the Scandinavian and Baltic countries have ENS in the altered scenarios. From the three scenarios provided by the TYNDP, National Trends has the least ENS while Distributed Energy has the most.

\begin{figure}[!htpb]
    \centering
    \includegraphics[width=\textwidth]{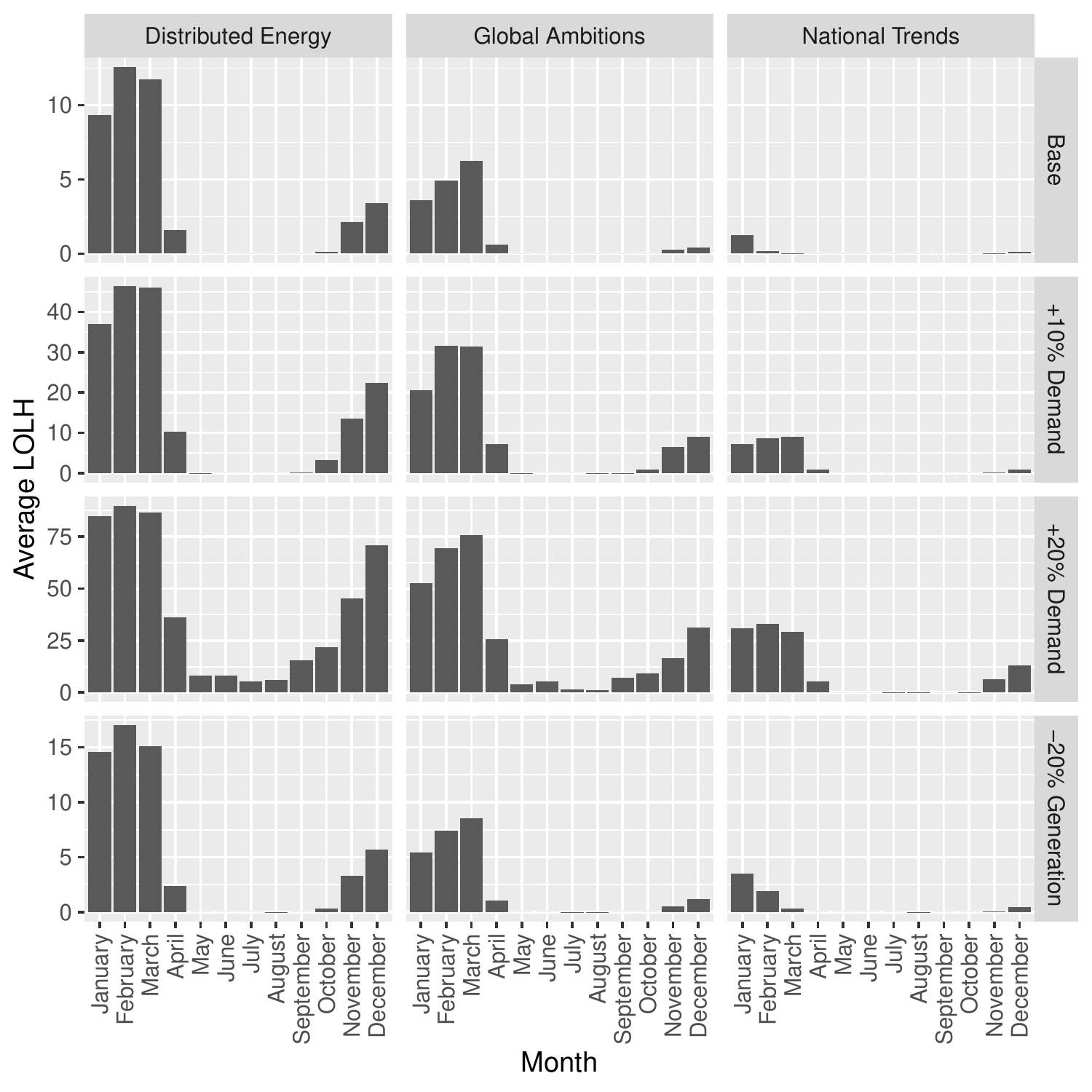}
    \caption{The average loss of load hours per year and bidding zone for each month and scenario based on the weather years 1982 to 2010.}
    \label{fig:LOLHinSeasons}
\end{figure}

The distribution of ENS events throughout the year is given in Figure~\ref{fig:LOLHinSeasons} for all twelve scenarios, showing that most ENS occurs in the winter months. As the ENS between bidding zones can differ in orders of magnitude, the total hours of loss of load expected (LOLH) is shown instead. When the demand increases, inevitably there would be some ENS in the non-winter months as can be seen in the third row. However, most LOLH and ENS occur in the winter time period, from December to March. As the meteorological variability in this winter period can be identified by using the definition of the weather regimes, a classification of the atmospheric state, we can analyse the relation between these regimes and the ENS events.

\begin{figure}[!htpb]
    \begin{subfigure}[t]{0.48\textwidth}
    \includegraphics[width=\textwidth]{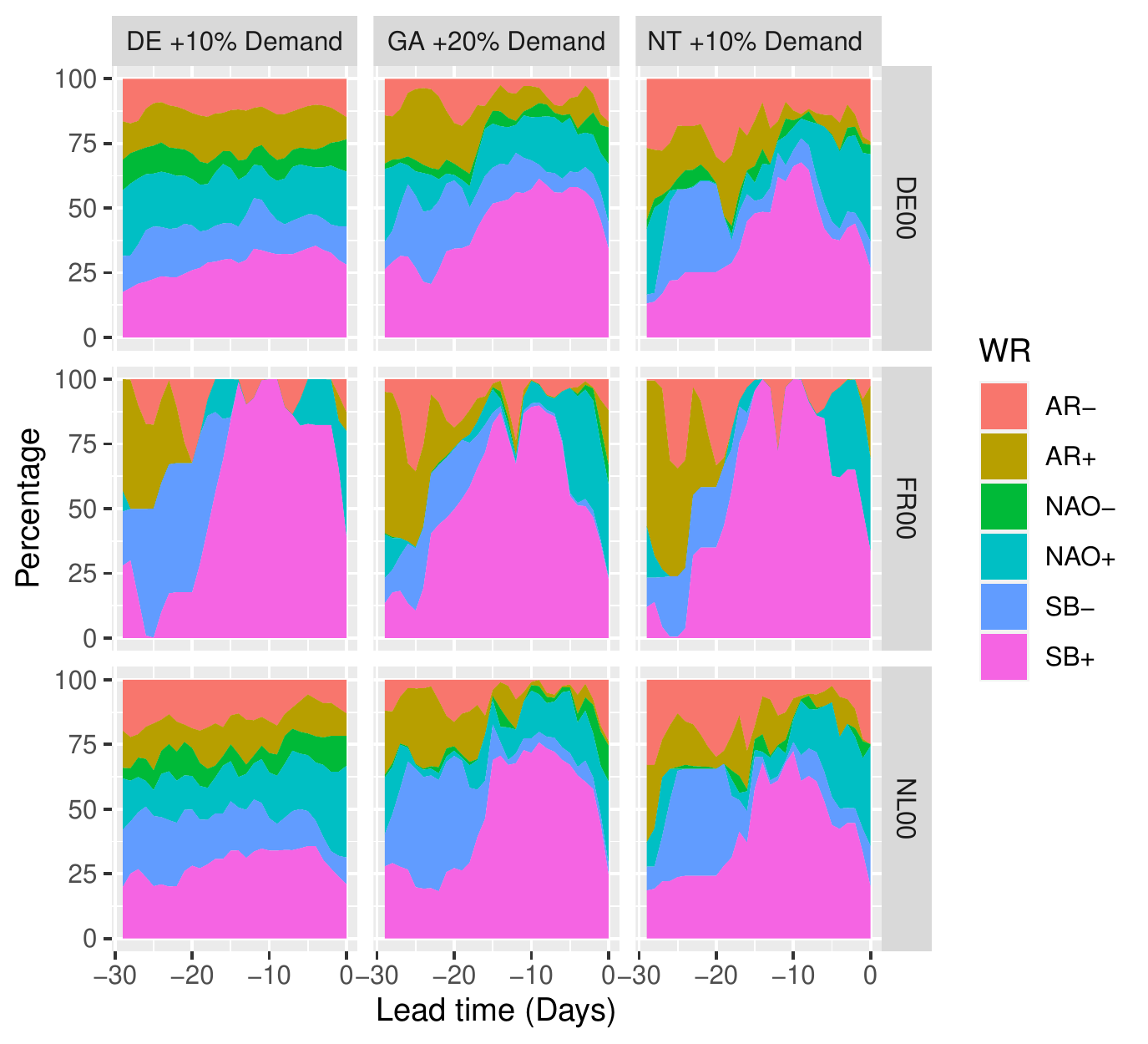}
    \caption{Central European bidding zones}
    \label{fig:WRPercentage1}
    \end{subfigure}
    \begin{subfigure}[t]{0.48\textwidth}
    \includegraphics[width=\textwidth]{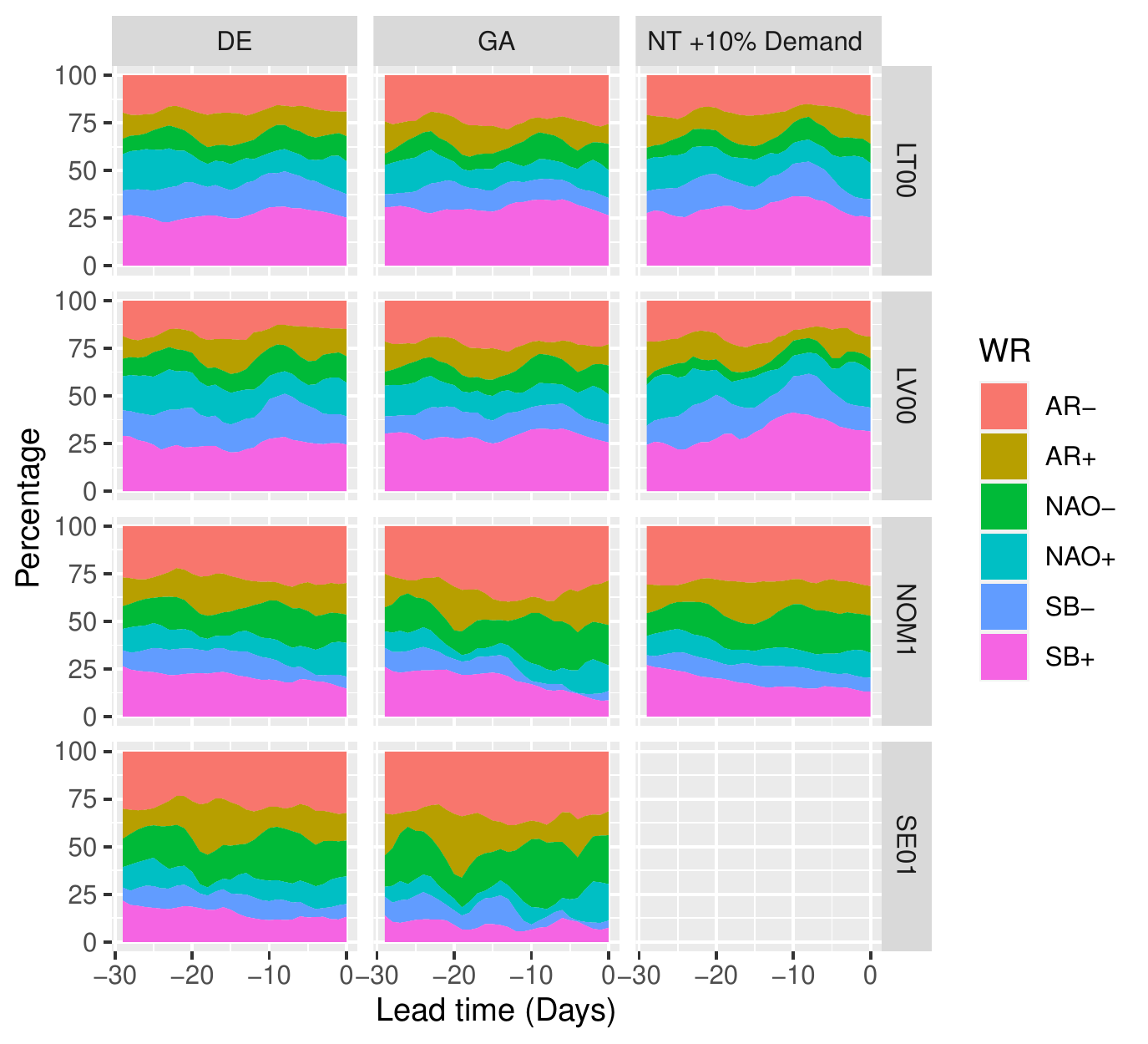}
    \caption{Scandinavian and Baltic bidding zones}
    \label{fig:WRPercentage2}
    \end{subfigure}
    \caption{The distribution of daily weather regimes (WR) occurrence in the 30 days before an ENS event for a subset of scenarios and bidding zones based on the weather years 1982 to 2010. }
    \label{fig:WRPercentage}
\end{figure}

To investigate whether ENS events are not only caused by a single atmospheric state, but also by a specific sequence of atmospheric states, the weather regime occurrence preceding an ENS event is shown in Figure~\ref{fig:WRPercentage}. We go back as far as 30 days before the ENS event to capture the impact of longer persisting weather regimes (see also Table~\ref{tab:WRTable}). We observe that three weather regimes are most prevalent in this 30 day period: the positive state of Scandinavian Blocking (SB+) and the North Atlantic Oscillation (NAO+) weather regimes for the typical central European regions (Figure~\ref{fig:WRPercentage1}), and the negative state of the Atlantic Ridge (AR-) regime for the Scandinavian and to a lesser degree at the Baltic zones (Figure~\ref{fig:WRPercentage2}). In the northern European zones this behaviour is already detected in the original scenarios. However, for central  European zones this behaviour is only detected during more challenging scenarios since those regions have almost no ENS in less demanding scenarios.

For the Netherlands in the scenarios DE+10\%, GA+20\% and NT+10\%, presented in Figure~\ref{fig:WRPercentage1}, the SB+ regime is present during 21.0\%, 24.9\% and 20.1\% of the ENS events, respectively. While its average occurrence is 15.0\% during the full period analysed (1982-2010). In the 30 days prior to an ENS event the prevalence of SB+ is even stronger for the Netherlands with 28.7\%, 47.0\% and 40.6\% of the time, respectively. 

Similar behaviour is seen in most central European bidding zones, and while its absolute statistics differ between scenarios, it does not seem to depend on a specific scenario. Giving the fact that SB+ is the least frequent weather regime in our data set (Table~\ref{tab:WRTable}), it is a clear signal that the SB+ weather regime is more likely to result in critical situations for central Europe.  

\subsection{Sequence of Weather Regimes}\label{sec:resultsSequence}
The central European regions mostly have a NAO+ or SB+ weather regime on the day of the ENS event (Figure~\ref{fig:WRPercentage1}). In addition, even without the SB+ weather regime during the ENS event, this weather regime is prevalent 10 days before the ENS event. This suggests that a specific sequence or precedence of a weather regime could cause ENS. 

\begin{figure}[!htpb]
    \centering
    \includegraphics[width=\textwidth]{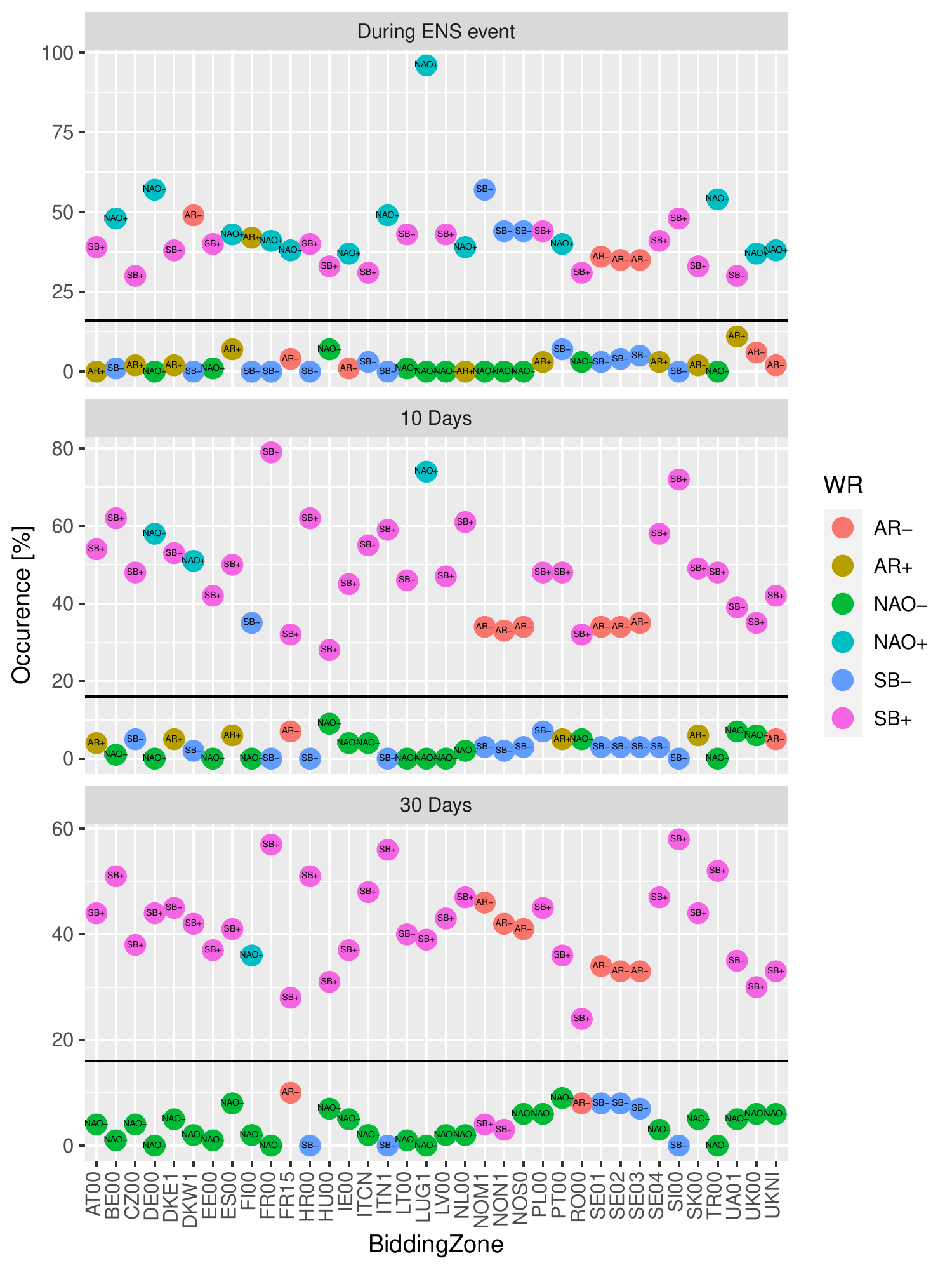}
    \caption{The occurrence of the two weather regimes that are present the most and the least during the day of the ENS event, and in the 10 and 30 days preceding it. Only the bidding zones with at least 50 ENS events across all scenarios and weather years based on 1982 to 2010 are presented. Naming convention are provided in Appendix B.}
    \label{fig:HighLowWr}
\end{figure}

To identify whether specific weather regimes are more prevalent in the period before ENS events, we assessed their occurrence during an ENS event, and 10 and 20 days before (see \ref{app:WRpriorENS}). We observe that ENS take place most of the time during the SB+ or NAO+ weather regimes across all scenarios (see Figure~\ref{fig:HighLowWr}), and 10 or 30 days before the ENS events the weather is mostly in a SB+ regime. However, for the bidding zones in Norway and Sweden the AR- weather regime occurs slightly more in the 10 and 30 days before an ENS event. During or 10 days before an ENS event, no weather regime is clearly occurring the least. However, 30 days before an ENS event, NAO- is the least present. 

\begin{figure}[!htpb]
    \centering
    \includegraphics[width=\textwidth]{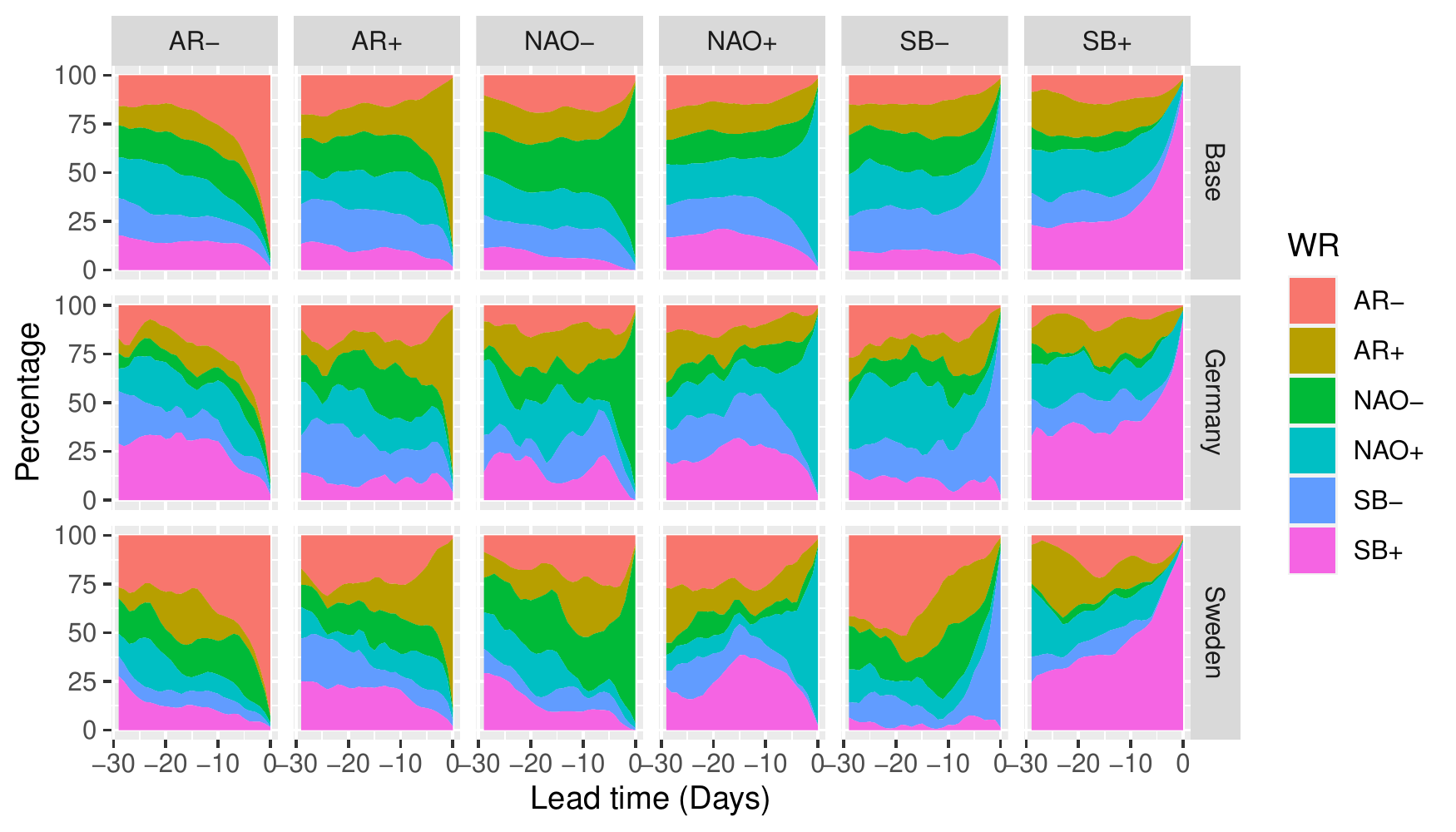}
    \caption{The daily relative occurrence of weather regimes (WR) before an ENS event grouped by the weather regime on the day of the ENS event in the analysed period (1982-2010). Base shows the normal precedence of weather regimes prior to a specific weather regime at day 0. For clarity, only Germany (DE00) and Sweden (SE01) are shown for the Distributed Energy +20\% demand scenario.}
    \label{fig:WRperWR}
\end{figure}

The occurrence of a possible specific sequence of weather regimes leading to ENS is shown in Figure~\ref{fig:WRperWR}. For central European bidding zones, represented by Germany, there is a significant occurrence of SB+ in the period leading to an ENS event that happens during the weather regimes AR-, NAO+ or SB+. We observe a peak in the presence of SB+ at 10+ days prior to the ENS event. This is not something that is already present in the normal precedence of these weather regimes.

For Scandinavian bidding zones, represented by Sweden in Figure~\ref{fig:WRperWR}, there is a significant occurrence of SB+ in the period leading to an ENS event that coincides with a NAO+ or SB+ weather regime. However, unlike the central European region this is not observed for ENS events during the AR- weather regime, which is most prominently associated with ENS events in northern Europe. In addition, we can see that the AR- regime is strongly present 20+ days prior to ENS events during SB- weather regime. 

\subsection{Driving factors for Unserved Energy during Weather Regimes}\label{sec:resultsDriving}
That some weather regimes have a stronger link with ENS events can be expected as they are a way to analyse the meteorological variability at a synoptic scale, which influences the renewable electricity generation, hydro inflow and temperature, which in turn influences electricity demand. However, these weather regimes are defined over the whole of Europe, while their impact depends on the region considered. 

\begin{figure}
    \centering
    \includegraphics[width=\textwidth, clip=true, trim=4cm 0 4cm 0]{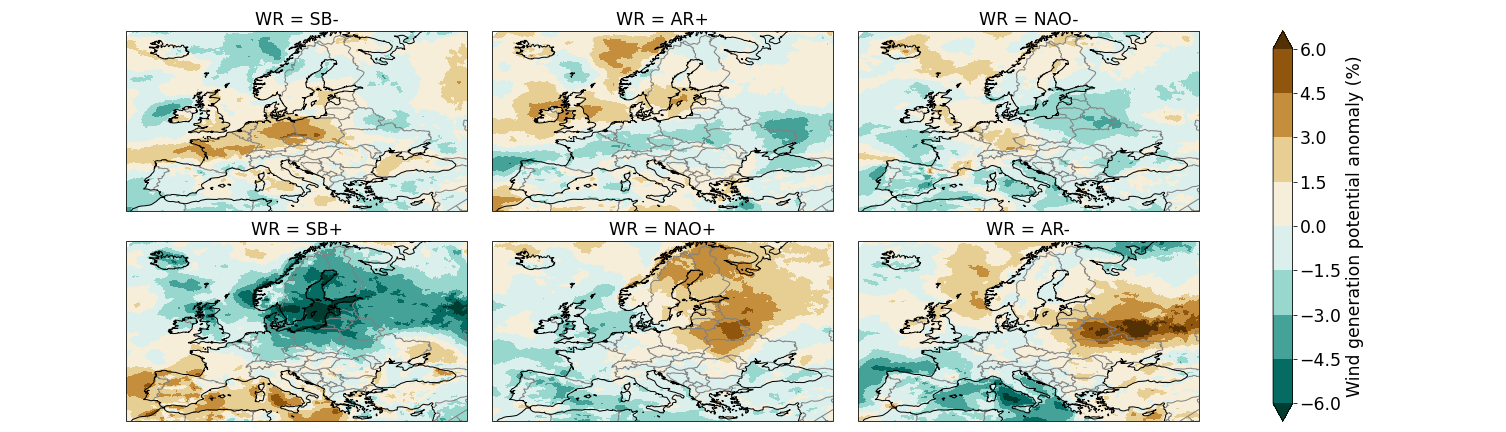}
    \caption{The average wind power generation anomaly from December to March (DJFM) in Europe for each weather regime with respect to the DJFM mean in the period 1982-2010.}
    \label{fig:WRwindCFDev}
\end{figure}

Based on the spatial changes of meteorological circumstances during a specific weather regime, we can assess its potential impact on the energy system. For instance, on average during the SB+ weather regime a decrease in wind power potential is observed in parts of central Europe with respect to the December to March mean (Figure~\ref{fig:WRwindCFDev}). This would likely impact the electricity system during this weather regime. Similarly, the NAO- weather regime is associated with small change in wind power potential over Europe. Additional Figures describing the anomaly over Europe in solar power potential and the meteorological drivers can be found in \ref{app:metWR}.

\begin{figure}
    \centering
    \includegraphics[width=\textwidth]{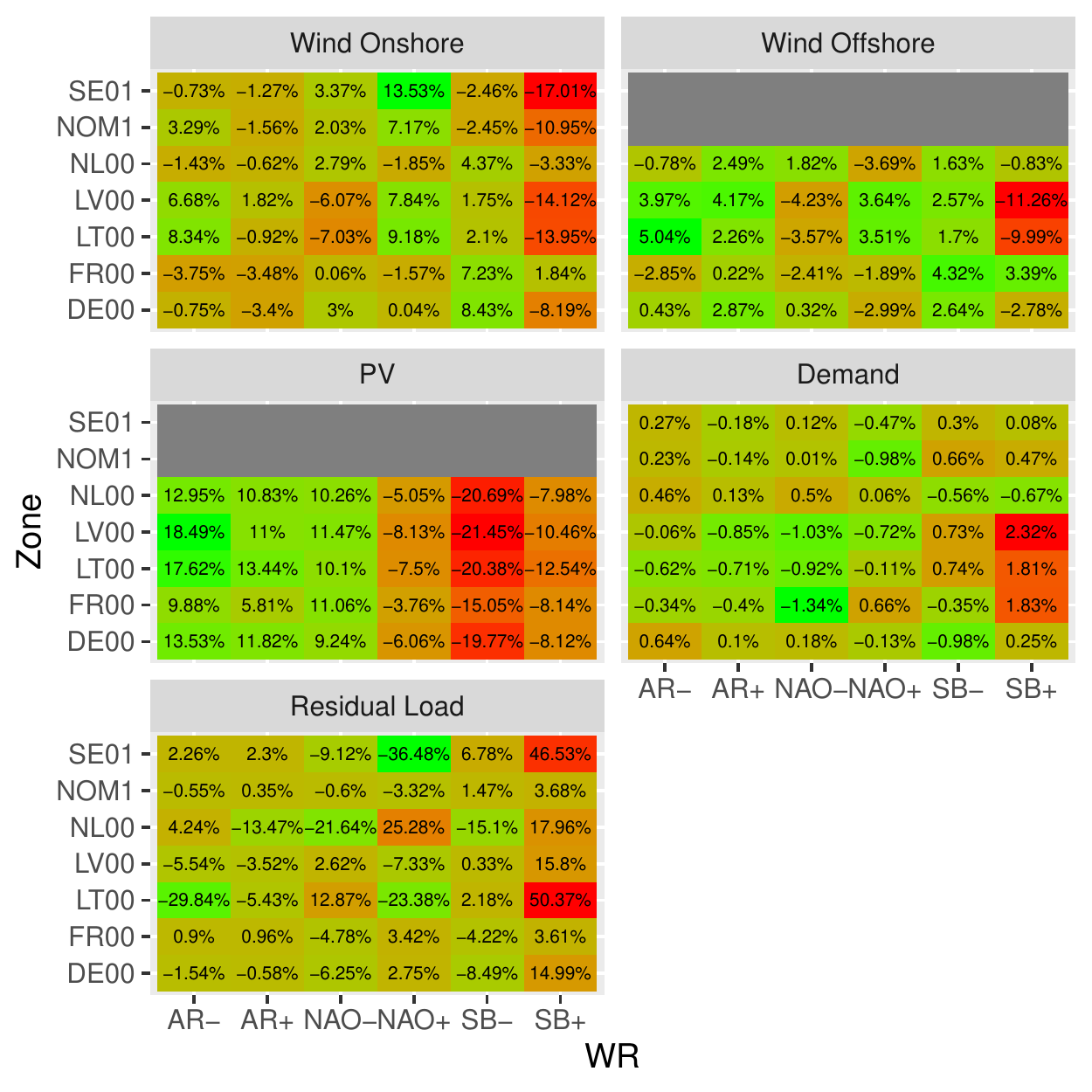}
    \caption{The average percentage anomaly, with respect to the December-March mean based on the weather years 1982-2010, in demand, solar photovoltaic (PV), residual load, onshore and offshore wind. The zones in the two typical regions for central Europe and northern Europe are shown. Note that residual load can be negative and the mean can be close to zero resulting in relative high changes.}
    \label{fig:ChangePerWR}
\end{figure}

Compared to the average anomaly in renewable energy generation potential, demand, and residual load in the winter period (see Figure~\ref{fig:ChangePerWR}), the SB+ weather regime is on average associated with higher demand and lower generation from solar photovoltaic systems, onshore and offshore wind. This behaviour is observed in both representative regions. Consequently, the residual load during a SB+ regime in these zones is much higher on average compared to other weather regimes, for the Latvian region (LT00) it is even 50.8\% higher. 

For some bidding zones another weather regime is associated with increased residual load. For instance, in the Netherlands (NL00), the NAO+ weather regime is associated with an even higher residual load on average ($25.9$\%) then under SB+ ($18.2$\%), while most ENS events are found during the SB+ weather regime. In addition, while the AR- regime shows a strong relation with ENS for the Scandinavian and to a lesser degree at the Baltic zones, the driving factors cannot be deduced based on the finding in Figure~\ref{fig:ChangePerWR}.

Apparently, the weather regime is not the only driving factor for ENS events and part of it is associated with (the assumptions of) the specific technologies used or the energy system scenarios. In addition, although the European electricity grid is well interconnected between bidding zones, the relation between an ENS event and a specific weather regime can change from bidding zone to bidding zone. However, in general the SB+ and NAO+ weather regimes are associated with more ENS events compared to other weather regimes.

\subsection{Role of storage}
As we have shown in Section~\ref{sec:resultScenario}--\ref{sec:resultsDriving}, the SB+ weather regime is prevalent in many bidding zones during ENS events. This weather regime can persist for a prolonged period (Table~\ref{tab:WRTable}), and has a peak in relative occurrence in the 10+ days prior to an ENS event. The strong presence of the SB+ weather regime in the period prior to an ENS event could indicate that a build up or a specific sequence of weather regimes is needed for ENS to occur. This could indicate that storage plays a role, as it may be depleted during these weather regimes. 

\begin{figure}
    \centering
    \includegraphics[width=0.6\textwidth]{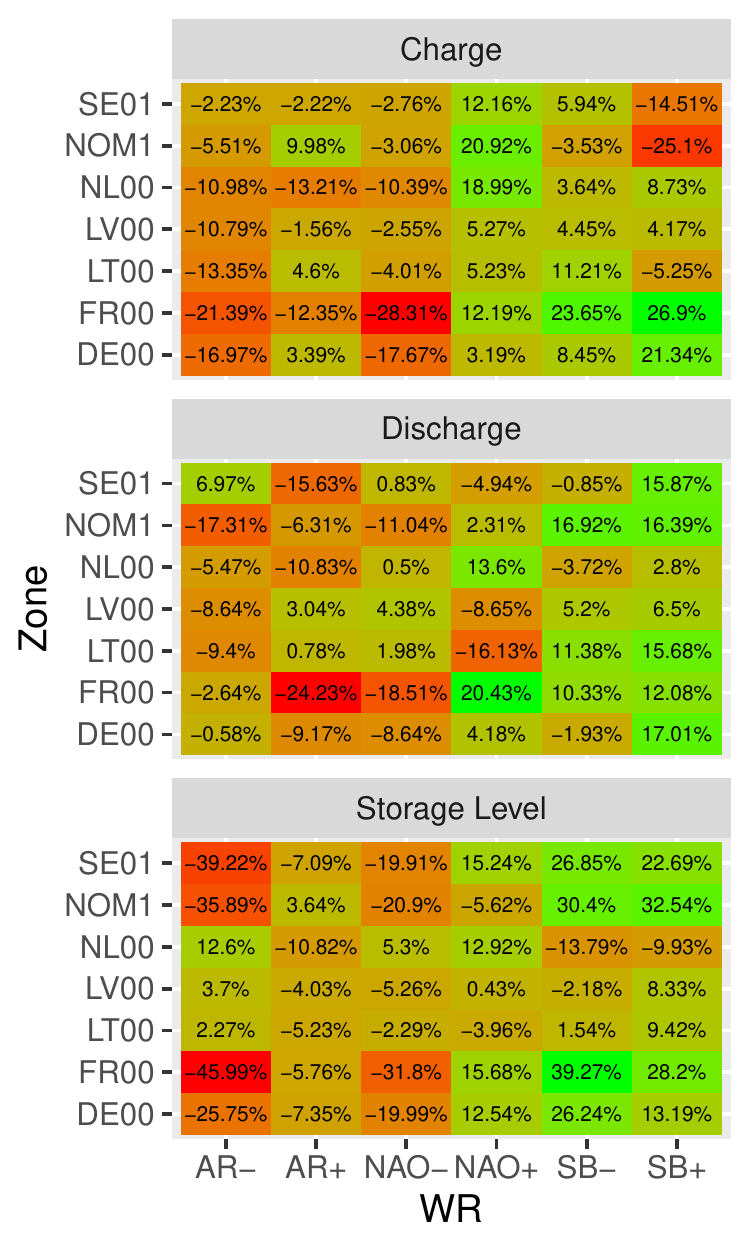}
    \caption{The average percentile anomaly per weather regimes in charge, discharge and storage level, compared to the December to March mean during the 1982--2010, for the total storage from batteries and (pumped) storage hydropower plant. The zones in the two typical regions for central Europe and northern Europe are shown.}
    \label{fig:ChangePerWRStorage}
\end{figure}

The evaluation of the average anomaly in storage level, the amount of charging and discharging during a specific weather regime (see Figure~\ref{fig:ChangePerWRStorage}) shows that the average storage levels is higher during the SB+ weather regime in all bidding zones except for the Netherlands (NL00). In addition, we observe an increase in the discharge (all regions) \emph{and} in the charging (central Europe) during these events. Similar behaviour is observed for the SB- weather regime and during the NAO+ regime for the Netherlands (NL00).

The earlier observation that SB+ is more challenging may be aligned with the observation in Figure~\ref{fig:ChangePerWRStorage} that storage is utilised more frequently during the SB+ weather regime to overcome these challenges. It should be noted that some regions, like Germany (DE00), France (FR00), Sweden (SE01) and Norway (NOM1), also show this behaviour to a lesser extend during the SB- and NAO+ weather regime. 

In addition, while the AR- regime shows a strong relation with ENS for the Scandinavian and to a lesser degree at the Baltic zones, the driving factors cannot be deduced based on the finding in Figure~\ref{fig:ChangePerWR}.

In addition, we observe that the both states of the AR and the NAO- weather regime are associated with low storage levels in the Scandinavian and some central European regions. Especially the AR- weather regime can be linked to depleting storage for Sweden (SE01), Norway (NOM1), and France (FR00), with storage levels at -39.2\%, -35.9\%, and -46.0\% respectively. As the AR- regime shows a strong relation with ENS for the Scandinavian and Baltic zones, storage is likely the driving factor in these zones.

\subsection{Validation of power system model formulation}
Model characteristics often included in the UCED problem such as ramping limits, minimum uptime, and binary commitment variables, are unnecessary when you are only interested in energy not served~\cite{wuijts2022modelchar}. This was validated by running multiple months with 4 different models of varying degree in detail and similar values of ENS where found (see \ref{appendix:PSM}). 

However, similar values of ENS do not imply that they occur at the same bidding zone or time~\cite{wuijts2022pitfalls}. Therefore, to make sure our results are robust under different model decisions, a more detailed model that includes generation cost (the Cost Model in \ref{appendix:PSM}) was simulated for analysed period of 1982-2010 for the Distributed Energy +10\% demand scenario. The Cost Model is optimised by the interior-point barrier method while the ENS Model is optimised by the dual simplex method.

\begin{figure}[!htp]
    \centering
    \includegraphics[width=0.6\textwidth]{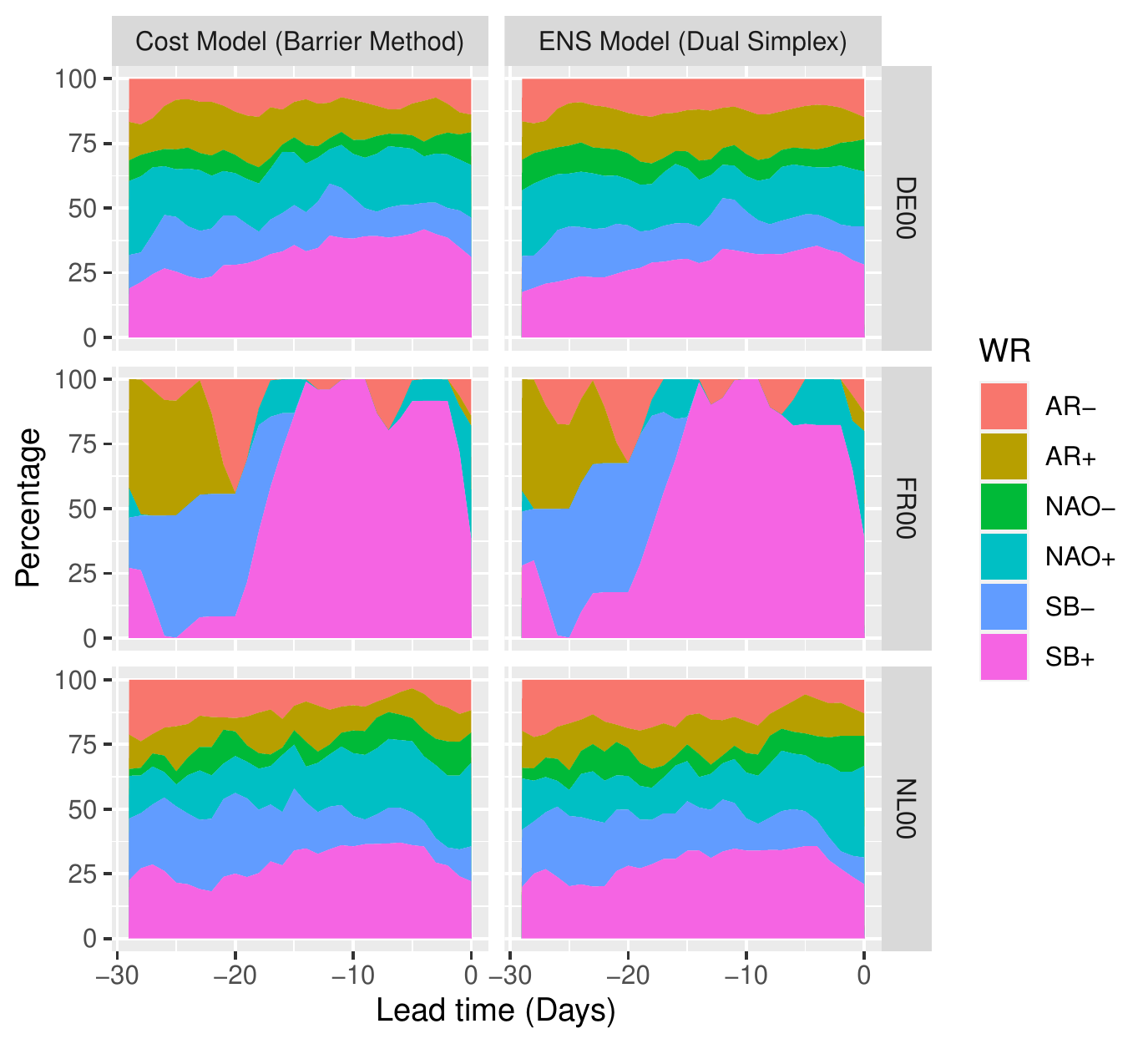}
    \caption{The distribution of daily weather regimes (WR) occurrence in the 30 days before an ENS event for the Cost and ENS Model formulations. Only the DE with 10\% extra demand scenario DE00, FR00 and NL00 bidding zones are shown based on the weather years 1982 to 2010.}
    \label{fig:PSMValidation}
\end{figure}

In line with the results shown in Figure~\ref{fig:WRPercentage} for the ENS Model, we observe that three weather regimes are most pronounced in the Cost Model (see Figure~\ref{fig:PSMValidation}). While the exact timing might be shifted slightly within a day, we observe extremely similar occurrence rates of the weather regimes prior to and during ENS events. Both model formulations identify the same critical moments, indicating that cost plays no role for possible occurrence of an ENS events.

\section{Limitations and discussion}
The relation found between ENS events and the weather regimes is in line with previous studies~\cite{Grams2017weather,vanderwiel2019extreme,vanderwiel2019regimes,Bloomfield2019regimes,Otero2022,Mockert2022arxiv,tedesco2023}. Although changes in the absolute values of risk differ between studies, an increase in risk during a winter time period high pressure system over Scandinavia or north-west Europe is seen. Where previous work has focused on the relation between weather regimes with renewable energy resource generation~\cite{Grams2017weather,ravestein2018vulnerability} or the residual load~\cite{vanderwiel2019regimes,Bloomfield2019regimes,Otero2022,Mockert2022arxiv,tedesco2023}, we analysed this relation between weather regimes and critical situations identified in full power system simulations. The link between specific weather regimes and ENS can be used to inform policy maker or grid operators in the choices they make. For instance, when in the long-term forecast a strong and persistent Scandinavian Blocking is observed, an early warning could be given to the grid operators to adjust their short-term planning for the likely reduced availability of the wind and solar resources. 

In our study, the analysed period is limited to 28 historical weather years due to the availability of data. This is slightly shorter than the 30 years normally used when looking at the impact of weather. By using a dataset with a longer period of consistent data, the relation between ENS and weather regimes can be better evaluated~\cite{Bloomfield2021nextgen}. At the same time, the analysis presented here covers a significantly larger number of weather years than used by Transmission System Operators (TSO) in their European resource adequacy assessments~\cite{ERAA2021}, and ten year network development plan (TYNDP) scenario assessment~\cite{entso2020tyndpguidelines}. In line with the recommendations made by \citeA{craig2022disconnect}, the development of the new open access version of the Pan-European Climate Database (PECD)~\cite{Dubus2022PECD} will provide the European TSOs and other parties with the option to use a longer consistent dataset, covering past and future projections, that allows for similar and more advanced assessments then were made here. 

The limited time-span of the period analysed imposed additional limitations on the study results, especially related to the uncertainties related to climate system. By only looking at a short historic period, we cannot adequately assess the aleatoric uncertainty, the year-to-year variability of weather, of our results. This interannual to multi-decadal variability of the state of the climate would preferably be addressed to obtain robust results~\cite{craig2022disconnect}. For instance, the combined wind and solar resource shows multidecadel variability of around $5$\%~\cite{Wohland2021}, which is a similar order of magnitude as the change observed during the challenging weather regimes (see Figure~\ref{fig:ChangePerWR}). 
In addition, the frequency, persistence and transition probabilities of a specific weather regime varies over decades~\cite{Dorrington2022} and is linked to the state of the El Nino Southern Oscilation (ENSO)~\cite{Falkena2022}. Consequently, by changing the specific decades analysed, the absolute values of the risk can differ. 

By only looking at historical weather years, no assessment of the impact of climate change could be made. As the TYNDP scenarios cover different states of future, it would be better if climate state of the future would be used to drive the weather dependent parts of the energy system~\cite{craig2022disconnect}. However, the climate state of the future is subject to three very strong sources of uncertainty that would need to be accounted for~\cite{Shepherd2019}. There is uncertainty in the specific pathway to the future and the emissions associated with this, these could be assessed by using multiple Representative Concentration Pathways (usually 3-4 RCP's are used). The epistemic uncertainty in the climate response to these emissions could be assessed by comparing the results from different climate models (depends on the region under consideration, usually 5-12 are used). And finally, the aleatoric uncertainty can be assessed by looking at a consecutive 30 year period. Assessing all three uncertainties for all TYNDP models used here (3 base or 12 with all adjustments) would require running at least $1350$ and upwards to $17280$ years through our Power Sytem Model. Even if a consistent dataset of all weather dependent variables could be created, this is currently not feasible due to the running time of our simulations (on a small cluster this takes 20-45 minutes for the ENS Model). Although there is no agreement on how exactly~\cite{craig2022disconnect}, a different approach should thus be used to assess the impact of climate change.

Because specific modelling choices in the UCED have an effect on the decision variables and outcomes of the simulation~\cite{wuijts2022modelchar}, model choices matter. For instance, the fixed storage level at start and end of the year might limit the use of storage in times of need, although similar assumptions are made in the \citeA{ERAA2021}. In addition, the analysed period we use in a single run ($8760$ hours, or one year) is already an improvement compared to the two-stage simulation used within the \citeA{ERAA2021}. To properly assess the impact of storage and its use during the ENS events a different start, and thus endpoint, of the simulated year would need to be considered. In addition, the level of storage defined in the scenarios might be subject to over planting with respect to the renewable energy resource capacity. As shown by \citeA{Livingston2020} the ability to balance the potential of renewable generation with storage has a limit in the order of hours to a few days and additional storage will likely rarely be utilised by renewables.

\section{Conclusion}
The aim of this study was to investigate the relationship between weather regimes and energy not served (ENS). For this we analysed twelve future European capacity scenarios based on 2020 Ten Year Network Development Plan from the European transmission system operators. For each of these scenarios we simulated an hourly power system model with weather dependent demand and renewable energy generation from 28 historic weather years. 

The different scenarios show slightly different results, but most ENS events occur in the period from December to March. We find that most bidding zones have a particular weather regime that causes the most ENS events. However, an ENS event can still occur in all regions during any weather regime but with a smaller probability. Different scenarios show some variation, but the weather regime associated most with an ENS event for a region is consistent across the scenarios analysed. 

For western European bidding zones, ENS events tend to coincide with the positive Scandinavian Blocking (SB+) and positive North Atlantic Oscillation (NAO+) weather regime. During the SB+ weather regime persistent cold and calm weather is observed, leading to an increased electricity demand in conjunction with a decreased wind and solar energy potential, which leads to high residual loads. While the NAO+ regime is associated with stronger westerly flow, and thus increased wind energy potential in Scandinavia, it is associated with ENS events in central Europe. This could be due to the observed prevalence of SB+ in the 10+ days prior to an ENS event in these regions. While storage utilisation is increased significantly, storage levels are not depleted on average during this weather regime. 

For Scandinavian and Baltic countries, the results indicate that the negative Atlantic Ridge (AR-) weather regime is more likely to be present during and leading up to ENS events. During the AR- weather regime the calm, sunny and cold weather in north-eastern Europe leads to a slightly increased demand in these regions. A significant decrease in charging of the storage system, and the of storage level of most regions is observed. This combination likely drives the ENS events in this region. 

To conclude, this article shows a clear correlation between specific weather regimes and unserved energy for some European countries. We found that the period preceding an ENS event is important. This indicates that for some ENS events a build-up is required, and this illustrates the variable nature of the energy system.


\appendix

\acknowledgments
This article is part of the Algorithmic Computing and Data Mining for Climate integrated Energy System Models (ACDC-ESM) project. This project aims to improve Energy System Models (ESM) by enhancing the algorithms underlining the ESM's and using methods to reduce the required input dataset size. This project brings together the scientific fields of big data analytics, advanced optimisation algorithms, meteorology, and energy system modelling to tackle these challenges. In this project Utrecht University works together with experts from RUG, KNMI and TenneT TSO B.V.. More information on the ACDC-ESM project can be found at \url{https://www.uu.nl/en/research/copernicus-institute-of-sustainable-development/algorithmic-computing-and-data-mining-for-climate-integrated-energy-system-models-acdc-esm}.

The content of this paper and the views expressed in it are solely the author’s responsibility, and do not necessarily reflect the views of TenneT TSO B.V..

The ACDC-ESM project and therefor, Rogier H. Wuijts and Laurens P. Stoop, received funding from the  Dutch Research Council (NWO) under grant number 647.003.005. To gain access to some datasources, Laurens P. Stoop is part of the IS-ENES3 project that has received funding from the European Union’s Horizon 2020 research and innovation programme under grant agreement No. 824084.  

 The hydro data was generated as part of 'Evaluating sediment Delivery Impacts on Reservoirs in changing climaTe and society across scales and sectors (DIRT-X)', this project and therefor, Jing hu, received funding from the European Research Area Network (ERA-NET) under grant number 438.19.902.  

The authors wish to thank dr. Ad J. Feelders and prof. dr. Ernst Worrell for their fruitful discussions and insights. 

The authors wish to thank dr. S.K.J. Falkena for providing the data of the daily weather regime definitions.

\section{CRediT author statement}
Conceptualisation: \emph{Laurens P. Stoop and Rogier H. Wuijts}, Formal Analysis: \emph{Rogier H. Wuijts, Laurens P. Stoop}, Data Curation: \emph{Laurens P. Stoop, Rogier H. Wuijts and Jing Hu}, Investigation and Writing - Original Draft: \emph{Rogier H. Wuijts and Laurens P. Stoop}, Writing - Review \& Editing: \emph{All listed authors}, Supervision: \emph{Marjan van den Akker, Gerard van der Schrier and Machteld van den Broek}. 
All authors (except Jing Hu) are part of the ACDC-ESM project. The ordering of the first two authors was decided by the order of their PhD defence; these co-first authors can prioritise their names as first authors when adding this paper’s reference to their r\'esum\'es.

\clearpage
\bibliography{agusample}

%
%
%
%
%

\clearpage
\section{Open Research}



The ERA5 data used for to obtain the RES potentials in this study are available at the Climate Data Store via \url{https://www.doi.org/10.24381/cds.adbb2d47} under the License to use Copernicus Products~\cite{hersbach2018era5,hersbach2020era5}. The implementation of the conversion models from meteorological variables to renewable energy generation potential used in the study are available at Github via \url{https://github.com/laurensstoop/CapacityFactor-CF} and \url{https://github.com/laurensstoop/EnergyVariables} with the MIT license. The Electricity Demand data used within the analysis is available at Zenodo via \url{https://www.doi.org/10.5281/zenodo.5780184} with the CC 4.0 license~\cite{defelice2021pecd}. The RES generation and demand data used in this study are available at ZENODO via \url{https://www.doi.org/10.5281/zenodo.7390479} with  CC BY-SA 4.0~\cite{laurens_p_stoop_2022_7390479}. 

All the capacity data used and information about their origin is included in an online dataset at \url{https://github.com/rogierhans/TYNDP2040ScenarioData}.

The Weather Regime data used for the categorisation of weather in the study are available upon request from dr. S.K.J. Falkena or the specific weather regime definition used for the period 1982-2010 used here from \url{https://github.com/laurensstoop/weatherregimes/data/processed/WR_k6_combo.csv} with the MIT license. The method describing their creation is presented in~\cite{falkena2020regimes} and the original implemented code can be found on \url{https://github.com/SwindaKJ/Regimes_Public}.

A web-application to get the raw sub-basin hydro inflow data underlying our aggregation is can be found at SMHI HypeWeb via \url{https://hypeweb.smhi.se/explore-water/historical-data/europe-time-series} with the CC BY-SA 4.0 license~\cite{Donnelly2015}. 
The daily hydro inflow data aggregated to country level used for the availability of the hydropower plants, run-of-river hydropower plants and pumped storage hydropower plants in the study will become available at Zenodo via \url{https://www.doi.org/10.5281/zenodo.7766457} with the CC BY-SA 4.0 license. The data archiving for this is underway, but the current file structure and size, single file of 50GB, is not usable on most systems.


\clearpage
\section{Region definition and naming convention}\label{app:regioncodes}
\begin{figure}
    \centering
    \includegraphics[width=\textwidth, clip=true, trim=1.5cm 1cm 1.5cm 0]{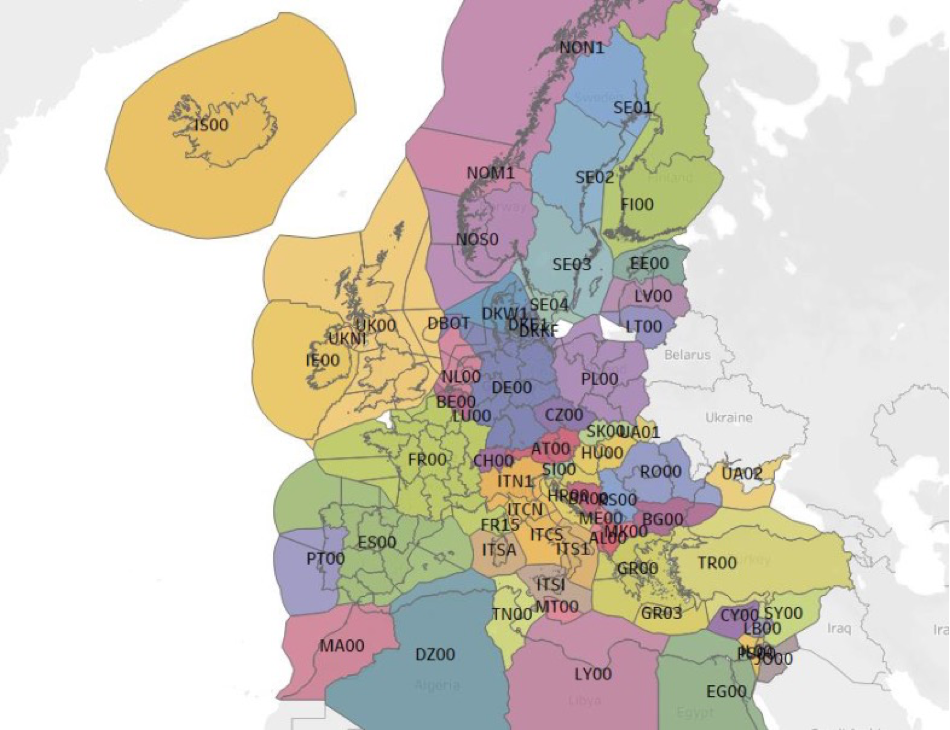}
    \caption{Location of the bidding zones used in this study. Figure provided by ENTSO-E. }
    \label{fig:BiddingzoneLocation}
\end{figure}

The bidding zone codes of the bidding zones used is shown in Table~\ref{tab:BiddingzoneCodes}, and their spatial location in Figure~\ref{fig:BiddingzoneLocation}. 
\begin{table}[ht]
\centering
\caption{Mapping between bidding zone codes and countries.}
\label{tab:BiddingzoneCodes}
\begin{tabular}{cc|cc|cc}
AL00 & Albania        & FR15 & France          & NL00 & Netherlands    \\
AT00 & Austria        & HR00 & Croatia         & NOM1 & Norway         \\
BA00 & Bosnia         & HU00 & Hungary         & NON1 & Norway         \\
BE00 & Belgium        & IE00 & Ireland         & NOS0 & Norway         \\
BG00 & Bulgaria       & ITCN & Italy           & PL00 & Poland         \\
CH00 & Switzerland    & ITCS & Italy           & PT00 & Portugal       \\
CY00 & Cyprus         & ITN1 & Italy           & RO00 & Romania        \\
CZ00 & Czech Republic & ITS1 & Italy           & RS00 & Serbia         \\
DE00 & Germany        & ITSA & Italy           & SE01 & Sweden         \\
DKE1 & Denmark        & ITSI & Italy           & SE02 & Sweden         \\
DKKF & Denmark        & LT00 & Lithuania       & SE03 & Sweden         \\
DKW1 & Denmark        & LUB1 & Luxemburg       & SE04 & Sweden         \\
EE00 & Estonia        & LUF1 & Luxemburg       & SI00 & Slovenia       \\
EL00 & Greece         & LUG1 & Luxemburg       & SK00 & Slovakia       \\
EL03 & Greece         & LV00 & Latvia          & TR00 & Turkey         \\
ES00 & Spain          & ME00 & Montenegro      & UA01 & Ukraine        \\
FI00 & Finland        & MK00 & North Macedonia & UK00 & United Kingdom \\
FR00 & France         & MT00 & Malta           & UKNI & United Kingdom
\end{tabular}
\end{table}

\clearpage
\section{Specified capacity of the main bidding zones used}\label{app:specificCAP} 
\begin{figure}[hb]
   \centering
   \includegraphics[width=0.9\textwidth]{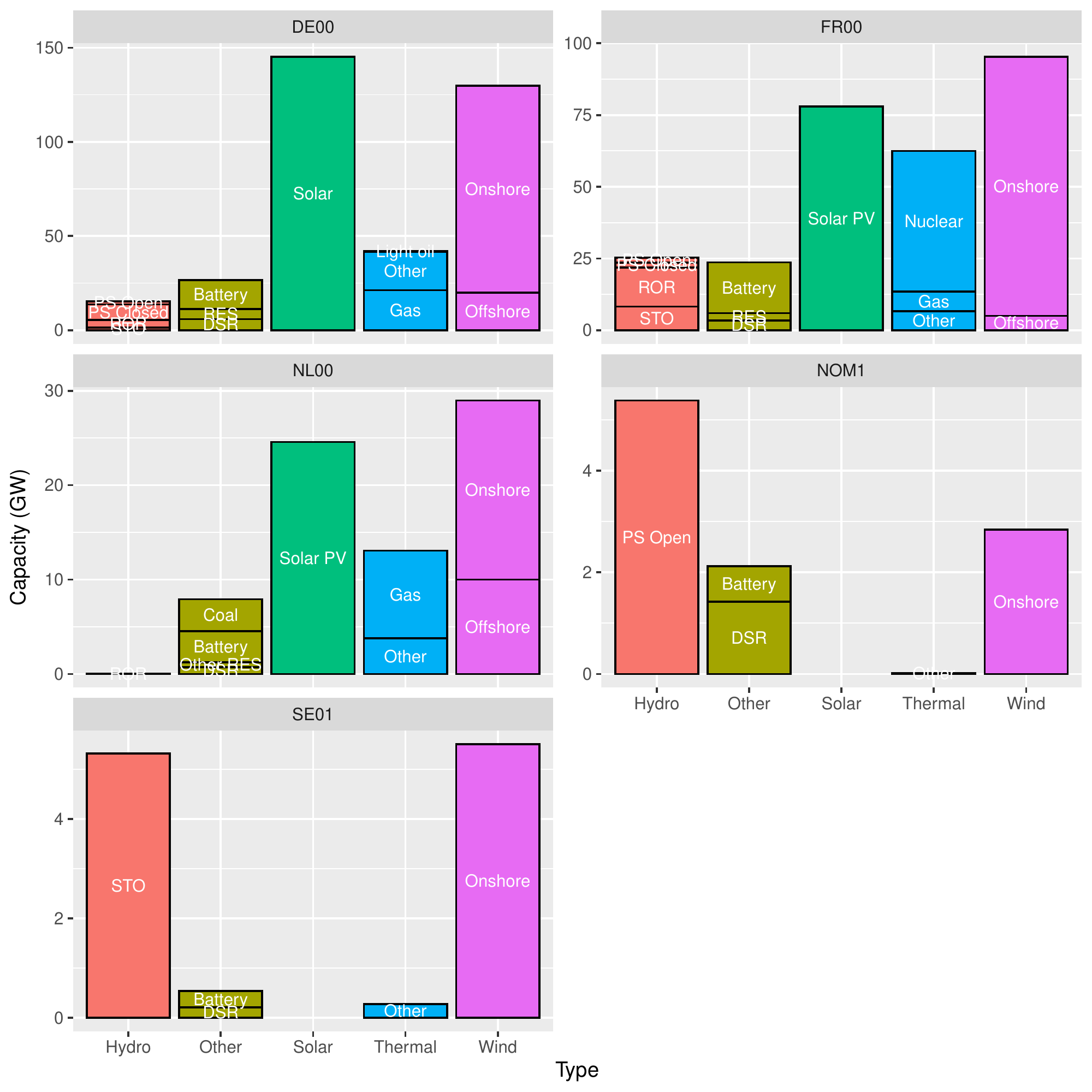}
   \caption{Installed capacity for a subset of bidding zones in the TYNDP Distributed Energy 2040 scenario is shown. The technologies listed are clustered according to their driving principle. }
   \label{fig:CapBZused}
\end{figure}

The specific installed capacities for the main bidding zones used in the analysis are shown in Figure \ref{fig:CapBZused}. The technologies are clustered based on their core principle in Hydro, Other, Solar, Thermal and Wind. The zones shown are the central European subset represented by Germany (\textit{DE00}), France (\textit{FR00}) and the Netherlands (\textit{NL00}), and the Scandinavian countries represented by the southern region of Norway (\textit{NOM1}) and the northern region of Sweden (\textit{SE01}). For the Scandinavian countries any region in Norway or Sweden could be used, for simplicity the first zone was thus chosen. 

\clearpage
\section{Unit Commitment Model}\label{appendix:PSM}
In this section, we define the Unit Commitment and Economic Dispatch (UCED) model that we use in this paper. Our UCED description is based on a Mixed Integer Linear Program (MILP) formulation with detailed thermal generators, Renewable Energy Sources (RES), storage, and transmission lines.
The decision variables $p_{gt}$, $p_{rt}$, $p_{st}$ are the generation of thermal generator ($g$), RES ($s$), and storage ($s$) unit at time step $t$. Other variables are added to narrow down the feasible state space of these variables. We use the well-known 3-bin formulation~\cite{knueven2020novel} which is given in:
\begin{align}
& \min \sum_{t \in T} \sum_{n \in N}VOLL \cdot ENS_{nt} + \sum_{g \in G} a_g u_{gt} + b_g p_{gt} +  v_{gt} cost_{start} \label{eq:chapter6:objective}\\
& s.t. \nonumber \\
& \production_{\unit \timeUnit} \geq \unitCommit_{\unit \timeUnit} \productionMin_\unit , \unit \in \unitSet, \timeUnit \in \timeSet \label{eq:chapter6:pmin}\\
&\production_{\unit \timeUnit} \leq \productionMax_g \unitCommit_{\unit \timeUnit} , \unit \in \unitSet, \timeUnit \in \timeSet \label{eq:chapter6:pmax}\\
&\sum_{i = \timeUnit - \minUpTime_g +1}^{\timeUnit} \decisionStart_{gi} \leq \unitCommit_{\unit \timeUnit} , \timeUnit \in \timeSet, \unit \in \unitSet \label{eq:chapter6:MDT1} \\ 
&\sum_{i = \timeUnit - \minDownTime_g +1}^{\timeUnit} \decisionStop_{gi} \leq 1- \unitCommit_{\unit \timeUnit} , \timeUnit \in \timeSet, \unit \in \unitSet \label{eq:chapter6:MDT2}\\ 
&\production_{\unit \timeUnit} - \production_{\unit \timeUnit-1} \leq (SU_g - \rampUp_g) \decisionStart_{g\timeUnit} + \rampUp_g \unitCommit_{g \timeUnit} , \timeUnit \geq 2, \unit \in \unitSet \label{eq:chapter6:tightRamp1} \\ 
&\production_{\unit \timeUnit-1} - \production_{\unit \timeUnit} \leq (SD_g - \rampDown_g) \decisionStop_{g\timeUnit} + \rampDown_g \unitCommit_{g\timeUnit-1} , \timeUnit \geq 2, \unit \in \unitSet \label{eq:chapter6:tightRamp2} \\
& p_{rt} \leq AF_{rt} \productionMax_{rt} , r \in R, t \in T \label{eq:chapter6:RES}\\
& 0 \leq \productionCharge_{\sUnit \timeUnit} \leq \maxCharge_\sUnit , \timeUnit \in \timeSet, s \in S\label{eq:chapter6:ChargeLimit} \\
& 0 \leq \productionDischarge_{\sUnit \timeUnit} \leq \maxDischarge_\sUnit , \timeUnit \in \timeSet, s \in S \label{eq:chapter6:DischargeLimit}\\ 
& p_{st} = \productionDischarge_{\sUnit \timeUnit} - \productionCharge_{\sUnit \timeUnit} , \timeUnit \in \timeSet, s \in S\label{eq:chapter6:StorageProduction} \\
& \minEnergy_\sUnit \leq \totalEnergy_{\sUnit \timeUnit} \leq \maxEnergy_\sUnit , \timeUnit \in \timeSet, s \in S \label{eq:chapter6:EnergyLimit}\\
& \totalEnergy_{\sUnit \timeUnit} = \totalEnergy_{\sUnit \timeUnit - 1} + \productionCharge_{\sUnit \timeUnit} * \effCharge_{\sUnit \timeUnit} - \frac{\productionDischarge_{\sUnit \timeUnit}}{\effDischarge_{\sUnit \timeUnit}} , \timeUnit \in \timeSet, s \in S \label{eq:chapter6:NetStorageProduction}\\
& \nodeInjection_{\node t } = \sum_{l=(n' \rightarrow n), \node' \in N} \flow_{lt} , \timeUnit \in \timeSet, \node \in \nodeSet \label{eq:chapter6:trans1}\\
&\underline{\flow_{l}} \leq \flow_{lt} \leq \overline{\flow_{l}} , l \in L, \timeUnit \in \timeSet \label{eq:chapter6:trans3}\\
& \sum_{\unit \in \unitSet_n} \production_{\unit \timeUnit} + \sum_{r \in R_n} p_{rt} + \sum_{s \in S_n}  p_{st} + \nodeInjection_{\node \timeUnit} = \demand_{\node \timeUnit} - ENS_{nt}, \timeUnit \in \timeSet, \node \in \nodeSet \label{eq:chapter6:nodalMarketClearing} \\
&\unitCommit_{gt} - \unitCommit_{gt-1} = \decisionStart_{gt} - \decisionStop_{gt} , \timeUnit \in \timeSet, \unit \in \unitSet \label{eq:chapter6:logic}\\
&  u_{gt},v_{gt},w_{gt} \in \{0,1\}, p_{gt},p_{rt},p_{st},pe_{st},inj_{nt},f_{lt} \in \mathbb{R} \label{eq:chapter6:domain}
\end{align}
(\ref{eq:chapter6:objective}) is the objective function of the UC consisting the system wide cost of Energy not Served (ENS) times the Value of Lost Load (VOLL) and the generation cost and start cost, $a_g$ is the constant cost and $b_g$ is the linear cost coefficient. Constraint (\ref{eq:chapter6:pmin}) and (\ref{eq:chapter6:pmax}) ensure the minimum and maximum production of generators. Constraint (\ref{eq:chapter6:MDT1}) and (\ref{eq:chapter6:MDT2}) ensure the minimum up and downtime of generators. Constraint (\ref{eq:chapter6:tightRamp1})  and (\ref{eq:chapter6:tightRamp2}) ensure the ramping limits of generators between time steps. Constraint (\ref{eq:chapter6:RES}) ensures that the RES production is lower than the availability at that hour. (\ref{eq:chapter6:ChargeLimit}), (\ref{eq:chapter6:DischargeLimit}) and (\ref{eq:chapter6:EnergyLimit}) ensure the charge, discharge, and energy storage limits for storage units. Equation (\ref{eq:chapter6:StorageProduction}) is the sum of charge and discharge i.e. the net storage production. Equation (\ref{eq:chapter6:NetStorageProduction}) describes the relation between the charge, discharge, and net power production of a storage unit.
 Equation (\ref{eq:chapter6:logic}) describes the logic between the binary commitment, start and stop variables of the generators.
 Equation (\ref{eq:chapter6:trans1}) describes the relation between the flow on transmission lines and the power injection at nodes. Constraint (\ref{eq:chapter6:trans3}) ensures flow limits on transmission lines.
 Equation (\ref{eq:chapter6:nodalMarketClearing}) ensures that the total generation meets the total demand at every node and time step.
At last, the commitment variables are binary while the generation are real numbers (\ref{eq:chapter6:domain}).

\section{Validating Assumptions}\label{app:validationPSM}
We assume that most model characteristics normally included in the unit commitment problem such as ramping limits, minimum uptime, and binary commitment variables, are unnecessary when you are only interested in energy not served~\cite{wuijts2022modelchar}. To validate this assumption, we ran four models with two different objective functions and two levels of detail to check whether this results in equivalent solutions. The first objective function minimises ENS directly, and the second one indirectly through the minimisation of total system costs in which ENS is heavenly penalised with a Value of Loss of Load (VOLL). 
Both types of minimisation's where applied to a full, and a simplified version of the UCED model. In the simplified version the ramping limits, minimum uptime, and binary commitment variables are omitted. We ran the four UCED models in chunks of 720 timesteps at the time for the weather years 1982--2010 and original 3 TYNDP scenarios.

We use the following naming convention to indicate the four different models used: 
\begin{itemize}
    \item Detailed ENS Model, including ramping limits, minimum up- and downtime and with binary variables. ENS minimisation.
    \item Detailed Cost Model, including ramping limits, minimum up- and downtime but  without binary variables. Cost minimisation.
    \item ENS Model, simplified model without binary variables. ENS minimisation.
    \item Cost Model, simplified model without binary variables. Cost minimisation.
\end{itemize}

The results, see Table~\ref{tab:ValidatingAssumpptions}, show that in all models the average ENS is almost the same. Moreover, the individual differences in all instances are smaller than 1~MWh, a negligible percentage of the total demand. The average computation time per weather year for these models differs significantly. As expected, a simplified model with ENS minimisation is significantly faster than running the model with more detail. Consequently, as the solution metrics that are relevant for our analysis (EENS) are the same, this is the preferred model to use as it allows us to run many power system configurations and weather years in order to get robust results.

\begin{table}[!htpb]
\centering
\caption{For the four different model formulations, the averaged total ENS in the system, the maximum difference in ENS of model run on a single weather compared to the average ENS of the same year and the average computation time of a year are shown. The average is calculated over all 720 hour chunks for the weather years 1982--2010. }
\label{tab:ValidatingAssumpptions}
\begin{tabular}{c|cccc}
Model & Detailed ENS     & Detailed Cost     & ENS     & Cost     \\ \hline
Average ENS (MWh) & 5246374.8 & 5246375.0 & 5246374.8 & 5246374.9 \\
Max difference (MWh) & 0.28  &    0.61   &  0.28   &  0.22\\
Avg. Computation time  (s) & 1051.9 &  593.6 &   52.7  &   190.9
\end{tabular}
\end{table}

\section{Method for determining photovoltaic panel generation potential}\label{appendix:Solar}
To obtain the solar photovoltaic potential generation, we follow the method as set out by \citeA{Jerez2015} Explicitly, the potential $PV_{pot}$ is calculated by formula \eqref{pvpot}.
\begin{align}
PV_{pot}&=P_R \frac{I}{I_{std}} \label{pvpot}
\end{align}
where $I$ is the short-wave downward radiation at the surface, $I_{std}$ is the incoming short-wave downward radiation under the standard test condition for solar photovoltaic cells ($I_{std} =1000$ W/m\ts{2}) and the performance ratio is given by $P_R$. 

The performance ratio can be modelled in a number of ways~\cite{Jerez2015}. Here, see Eq.~\eqref{pr}, we take into account the cooling effect of the wind on a solar panel cell temperature, which in turn is also influenced by the irradiance and the ambient air temperature, see Eq.~\eqref{tcell}. 
\begin{align}
P_R &= 1+\gamma \left(T_{cell} - T_{ref}\right) \label{pr}
\end{align}
where $\gamma=-0.5$ $\% \degree$C and the $T_{ref}=25$ $\degree$C is the standard test condition temperature for photovoltaic cells. The cell temperature $T_{cell}$ is modeled by formula \eqref{tcell}.
\begin{align}
T_{cell}&=c_1 +c_2 T+c_3 I+c_4 V\label{tcell}
\end{align}
where $T$ is the air temperature around the cell, $I$ the short-wave downward irradiance on the cell and $V$ the wind around the cell. The constants $c_1$ to $c_4$ have been determined by \citeA{tamizhmani2003} to be $c_1 =4.3$ $\degree$C, $c_2=0.943$, $c_3=0.028$ $\degree$C m\ts{2} W\ts{-1}  and $c_4= -1.528$ $\degree$C s m\ts{-1}.

\section{Method for determining wind turbine generation potential}\label{appendix:Wind}
To convert windspeeds to wind turbine generation potential we use an adjusted version of the power curve method from~\cite{Jerez2015}. We made three adjustments to this model. First, we reduced the effective capacity factor ($CF_e$) with $5\%$ to $95\%$ to represent the wake losses in large scale wind-farms. Secondly, we introduce a linear decay in the capacity factor at high wind speeds to more accurately represent high windspeed operational conditions. The third change was that we tuned the power curve regimes. Equation~\eqref{windpot} gives the capacity factor for wind turbines ($CF_{wind}$) used in this study. 
\begin{align}
CF_{wind}(t) &= CF_e \times
\begin{cases}
0 & \mbox{if} \qquad V(t)<V_{CI},\\ 
\frac{V(t)^3 - V_{CI}^3}{V_R^3-V_{CI}^3} & \mbox{if} \qquad V_{CI}\leq V(t)<V_R,\\ 
1  & \mbox{if} \qquad V_R \leq V(t)<V_{D},\\ 
\frac{V_{CO} - V(t)}{V_{CO}-V_{D}}  & \mbox{if} \qquad V_{D}\leq V(t)<V_{CO},\\ 
0 & \mbox{if} \qquad V(t)\geq V_{CO}.
\end{cases} \label{windpot}
\end{align}
Here $V(t)$ is the wind speed at the height of the wind turbine and the power curve regimes are given by the cut-in ($V_{CI}$=3 m/s), rated ($V_{R}$= 11 m/s), decay ($V_{D}$= 20m/s) and cut-out ($V_{CO}$= 25m/s) wind speed. The windspeed provided by ERA5 (at 100 meter) did not match the hub height for the wind turbines used in the TYNDP scenario. Using the wind profile power law we scaled the windspeed to 120 and 150 meters for on- and offshore turbines. The surface roughness was set to a constant value for both onshore ($\alpha=0.143$) and offshore regions ($\alpha=0.11$) in line with the reported values in \citeA{Hsu1994}.

\section{Method for determining Hydropower generation potential}\label{app:hydropowermodel}
The  hydro inflow data are based on historical river runoff reanalysis data simulated by the E-HYPE model~\cite{donnelly2016using}. E-HYPE is a pan-European model developed by The Swedish Meteorological and Hydrological Institute (SMHI), which describes hydrological processes including flow paths at the subbasin level. A subbasin in the context of hydrology is the region from which all surface run-off flows through to a particular point, this is generally a the collection of upstream streams, rivers and lakes. 

E-hype only provides the time series of daily river runoff (in m\textsuperscript{3}/s) entering the inlet of each European subbasin over 1980-2010. To match the operational resolution of the dispatch model, we linearly downscale the time series to hourly. By summing up runoff associated with the inlet subbasins of each country, we also obtain the country-level river runoff. 

The hydro inflow time series per country as inputs of the dispatch model is defined in this study as normalized energy inflows (per unit installed capacity of hydropower) embodied in the country-level river runoff. Therefore, it resembles an input capacity factor time series that can be extracted from water ($CFW_t$). The UCED model decides whether the energy inflows are actually used for electricity generation, stored, or spilled (in case the storage reservoir is already full). The hydro inflows $CFW_t$ at a given time $t$ is proportional to the instantaneous river runoff $Q_t$:
\begin{equation}
    CFW_t = \frac{Q_t}{Q_{avg}}  CFW_{avg}
\end{equation}
Where $Q_{avg}$ is the long-term average river runoff and $CFW_{avg}$ is the corresponding average energy inflow. However, the long-term average data of $CFW_avg$ are not available and cannot be calculated due to the lack of plant-level hydrological details such as hydraulic head; active storage volume. For practical reasons, we use the long-term capacity factors based on the actual electricity outputs of aggregated hydropower plants to replace $CFW_{avg}$. They can be calculated based on the average of reported  yearly capacity factors $CF_{avg}$~\cite{EUROSTAT2021}. The calculation would be ideally carried out for the entire 30-year period 1980-2010, but yearly capacity factors reported for many European countries are not readily available prior to 1990. Therefore, we calculated the ratio between the $CFW_avg$ and $Q_{avg}$ for the period 1990-2010 and used it as a scalar for the entire 30-year $Q_t$ time series.

In reality $CF_{avg}$ may be smaller than $CFW_{avg}$, because not all energy embedded in the river runoff is always converted into electricity due to spillage or non-energy usage. As a result, the values in the $CFW_t$ time series used as inputs in the UCED model may be underestimated.  

We explicitly consider three types of hydropower plants, namely storage hydropower plant (STO), run-of-river hydropower plant (ROR) and pumped storage hydropower plant (PHS). For modelling purposes, we need to estimate the specific maximum energy storage content (or specific storage size [$\frac{GWh}{MW}$]), for each type of hydropower. This is performed by deriving an EU-average specific energy storage content $\frac{GWh}{MW}$ per hydropower type based on an in-house database containing 207 large power plants. The derived specific energy storage content is calibrated to the present level of total storage size (220~TWh) of STO, RoR, and PHS together in Europe reported by~\cite{mennel2015hydropower}. The resulting specific energy storage contents for STO, RoR and PHS are 2.05, 0.43 and 0.18 $\frac{GWh}{MW}$. 

\section{Classification of weather into regimes}\label{app:weatherregimes}
To classify the European winter time period meteorological variability at the synoptic scale, a number of different criteria can be used. Although this classification is generally according to the anomaly of the geopotential height at the 500~hPa level, they have a strong relation with the variability at the surface~\cite{Grams2017weather,Thornton2017,Bloomfield2019regimes}. Recently the Grosswetterlagen method~\cite{baur1944kalender} has gained more traction due to the use of machine learning to classify the weather~\cite{neal2016flexible}. However, the large number of regimes used limits the interpretability of a single regime~\cite{neal2016flexible,falkena2020regimes}. For this reason, the use of the four European weather regimes based on a $k$-means clustering as set out in \citeA{michelangeli1995weather} is still prevalent in many impact studies (e.g. when considering the impact of the variability of weather on the renewable energy resource~\cite{vanderwiel2019regimes,Bloomfield2019regimes,Otero2022,tedesco2023}). On the other hand, when only four weather regimes are used to describe the full variability in surface weather, it is difficult to separate partially mixed signals that are present due to the course classification of the meteorological variability~\cite{Grams2017weather}. 

The use of four regimes to classify the winter time period weather in Europe is mostly due to the initial work by \citeA{michelangeli1995weather} on using $k$-means clustering on the first few empirical orthogonal functions of the geopotential height at the 500hPa level. While the granularity of the meteorological data has increased by 3-orders since then. Recently, \citeA{falkena2020regimes} revisited the identification of European weather regimes. They found that when the full field data is used, instead of only the first few empirical orthogonal functions, the optimal number of weather regimes is six when looking at the Bayesian Information Criterion. In addition, they show that by incorporating a weak persistence constraint instead of using a low-pass filter to stabilise the regime identification, the classification of a specific regime cluster is better and thus more defined.

\clearpage
\section{Summary of relative occurrence of weather regimes for all bidding zones}\label{app:WRpriorENS}
\begin{figure}[!hb]
    \centering
    \includegraphics[width=0.95\textwidth]{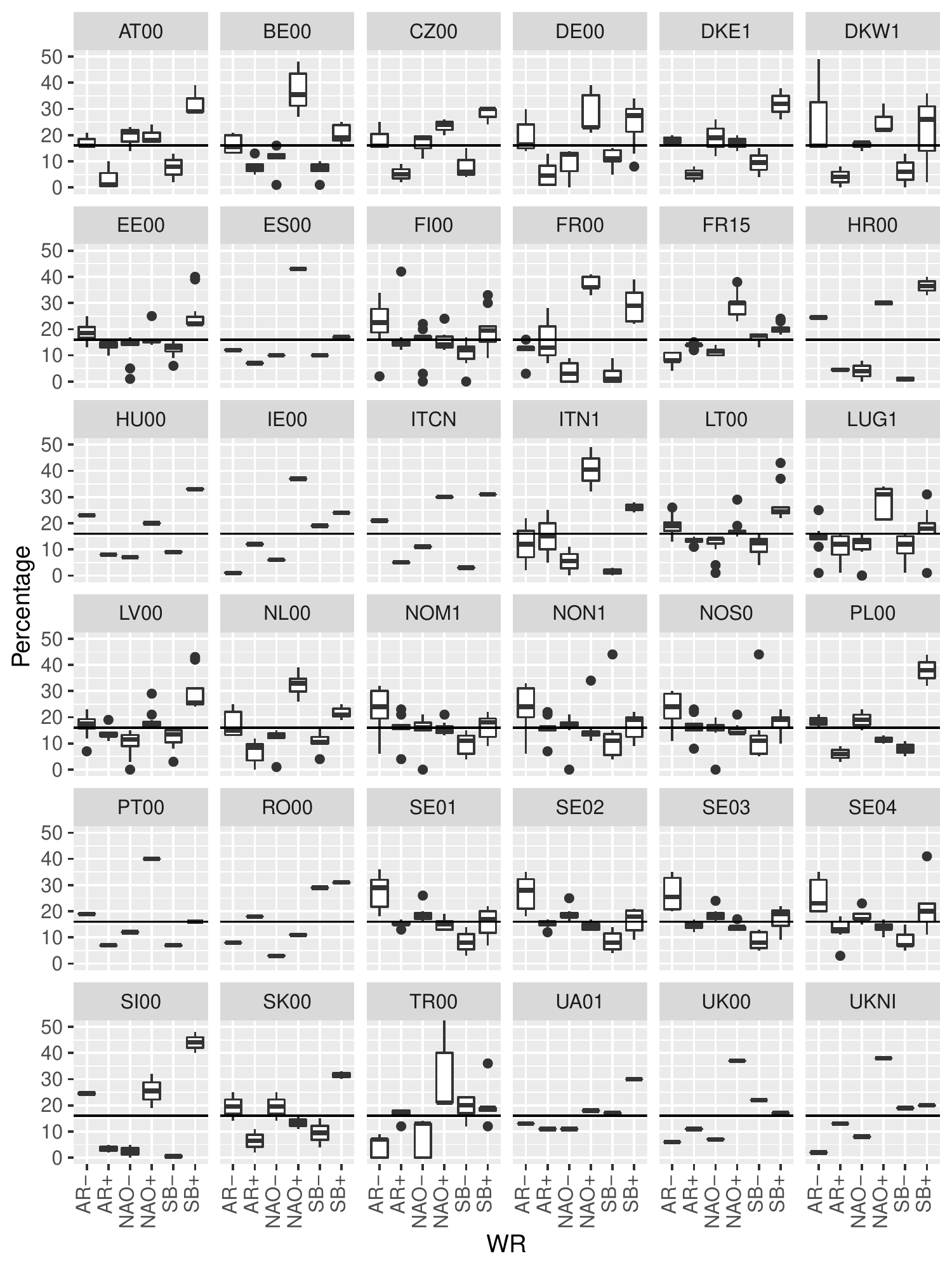}
    \caption{For each bidding zone and for all 12 scenarios the occurrence of weather regimes within a 1 day period before the ENS event.}
    \label{fig:Boxplot1Day}
\end{figure}

\begin{figure}[!ht]
    \centering
    \includegraphics[width=\textwidth]{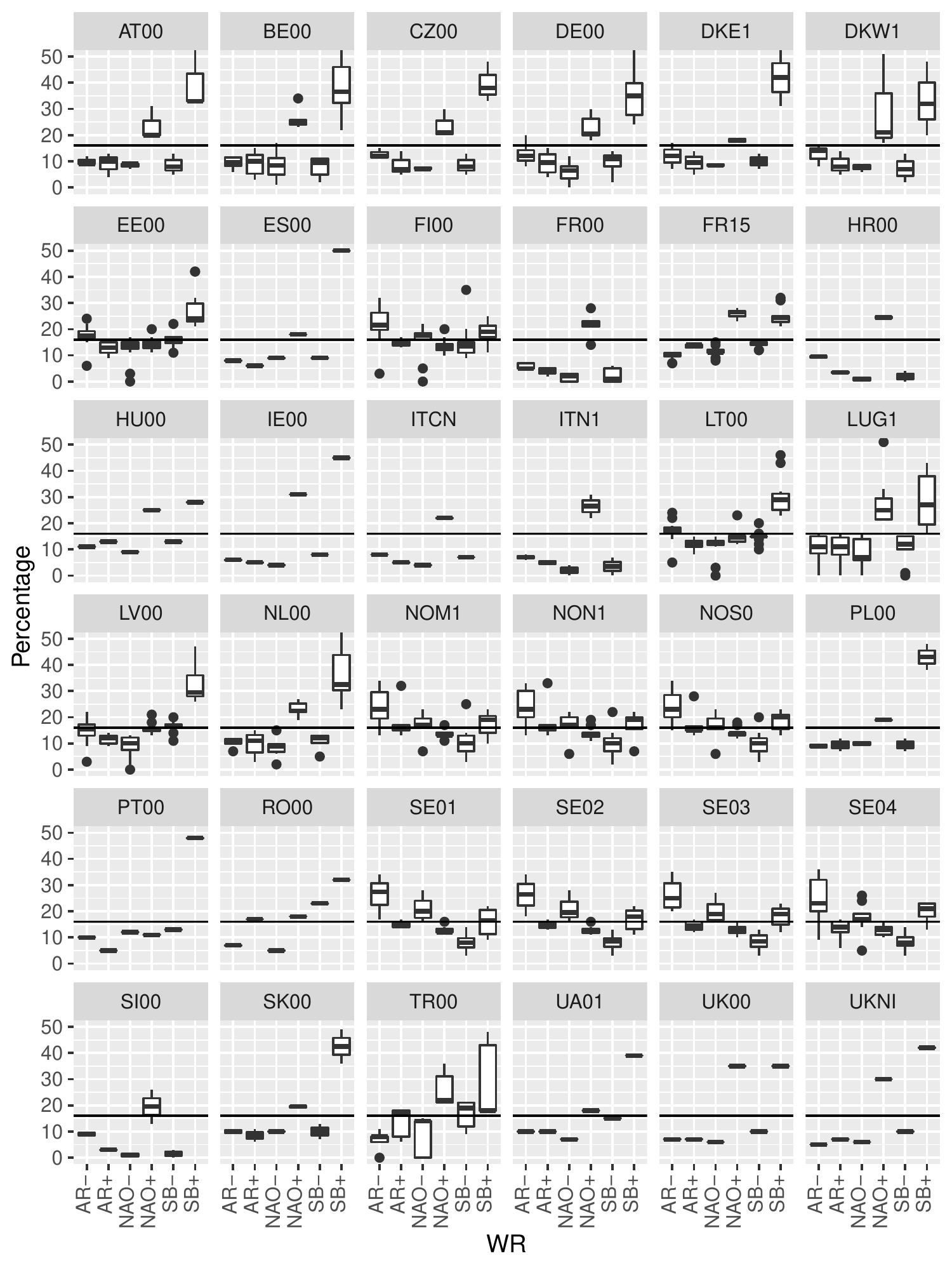}
    \caption{For each bidding zone and for all 12 scenarios the occurrence of weather regimes within a 10 day period before the ENS event.}
    \label{fig:Boxplot10Days}
\end{figure}

\begin{figure}[!ht]
    \centering
    \includegraphics[width=\textwidth]{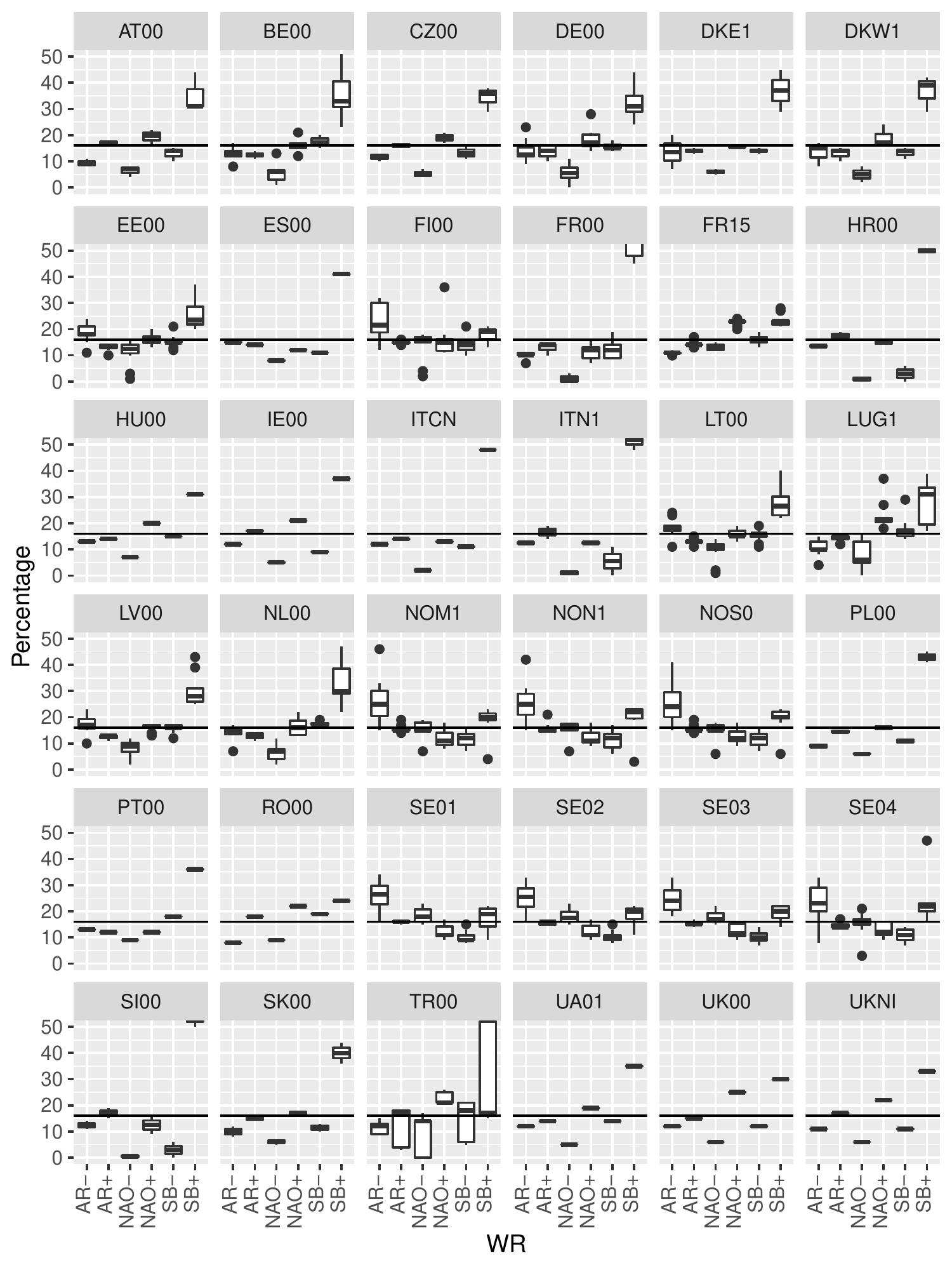}
    \caption{For each bidding zone and for all 12 scenarios the occurrence of weather regimes within a 30 day period before the ENS event.}
    \label{fig:Boxplot30Days}
\end{figure}

An overview of the relative occurrence of the weather regimes in the 1, 10 and 30 days prior to an ENS event for all bidding zones and scenarios is provided in figures \ref{fig:Boxplot1Day}, \ref{fig:Boxplot10Days} and \ref{fig:Boxplot30Days}. 

\clearpage
\section{Meteorological changes between different weather regimes}\label{app:metWR}
\begin{figure}[!h]
    \centering
    \includegraphics[width=1.3\textwidth,clip=true, trim=5cm 0 0 0 ]{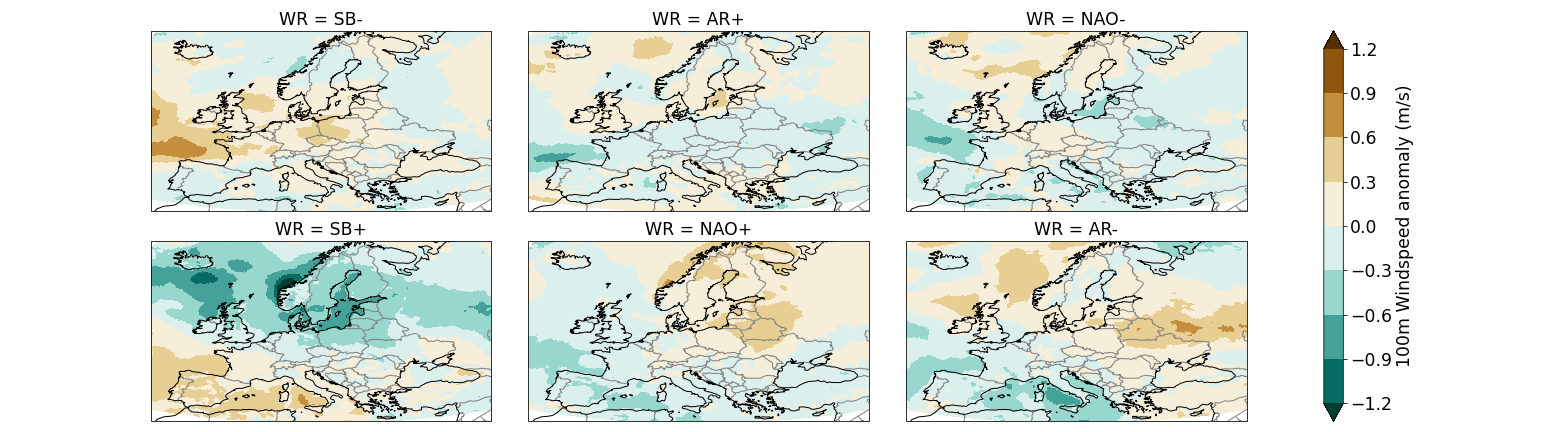}    
    \caption{The average 100 meter windspeed anomaly from December to March (DJFM) in Europe for each weather regime with respect to the DJFM mean over 1982-2010.}
    \label{fig:WRwindspeedDev}
\end{figure}

\begin{figure}[!hb]
    \centering
    \includegraphics[width=1.3\textwidth,clip=true, trim=5cm 0 0 0 ]{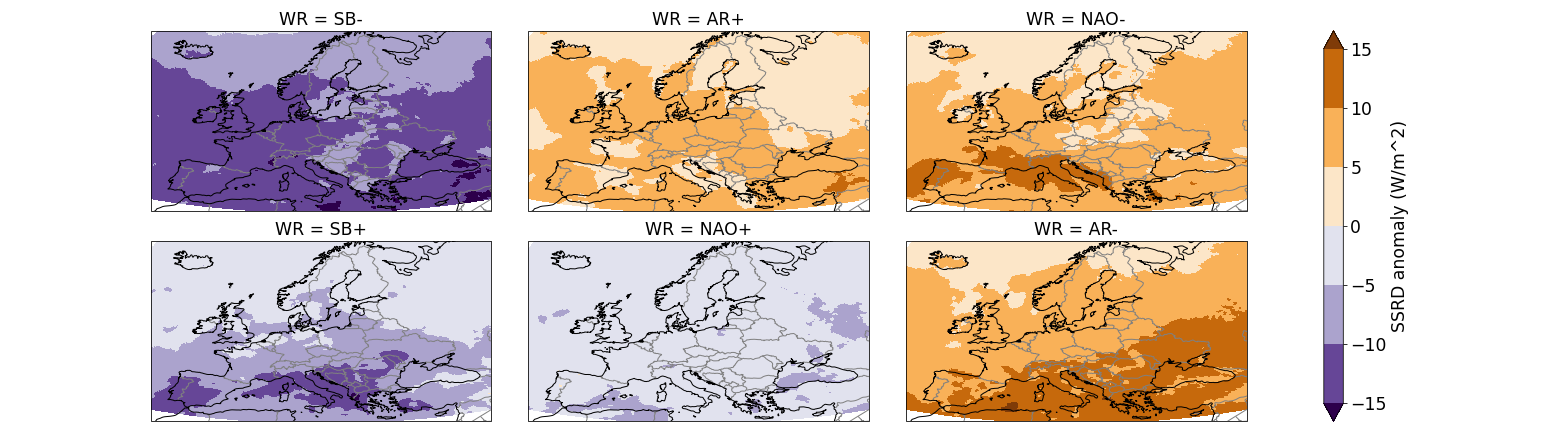}    
    \caption{The average solar irradiance anomaly from December to March (DJFM) in Europe for each weather regime with respect to the DJFM mean over 1982-2010.}
    \label{fig:WRSSRDDev}
\end{figure}

\begin{figure}[!ht]
    \centering
    \includegraphics[width=1.3\textwidth,clip=true, trim=5cm 0 0 0 ]{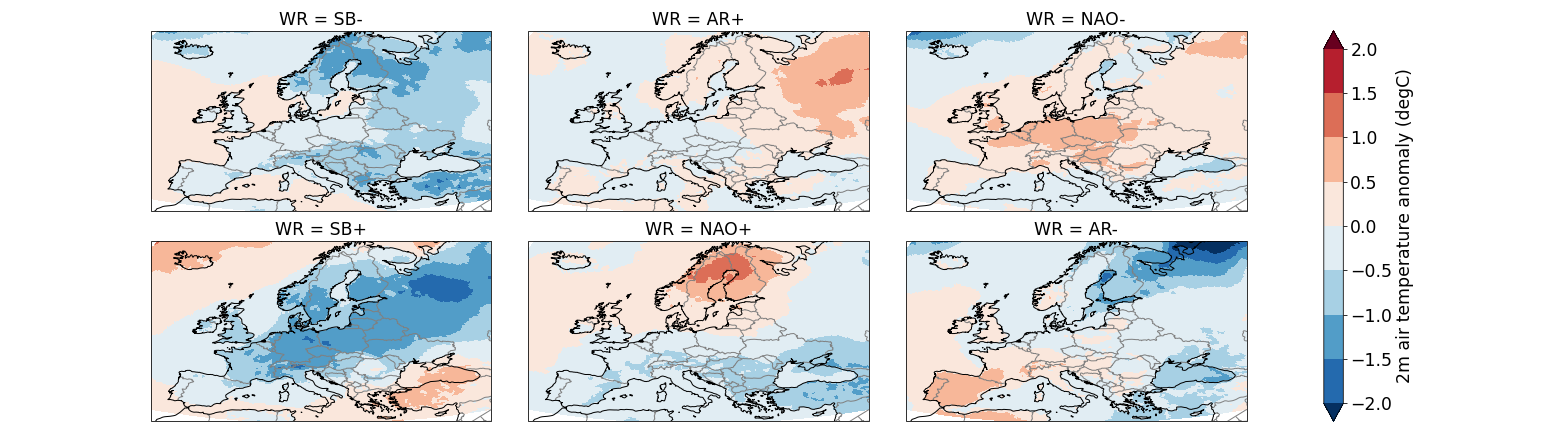}
    \caption{The average 2 meter air temperature anomaly from December to March (DJFM) in Europe for each weather regime with respect to the DJFM mean over 1982-2010.}
    \label{fig:T2mDev}
\end{figure}

\begin{figure}[!hb]
    \centering
    \includegraphics[width=1.3\textwidth,clip=true, trim=4.5cm 0 0 0 ]{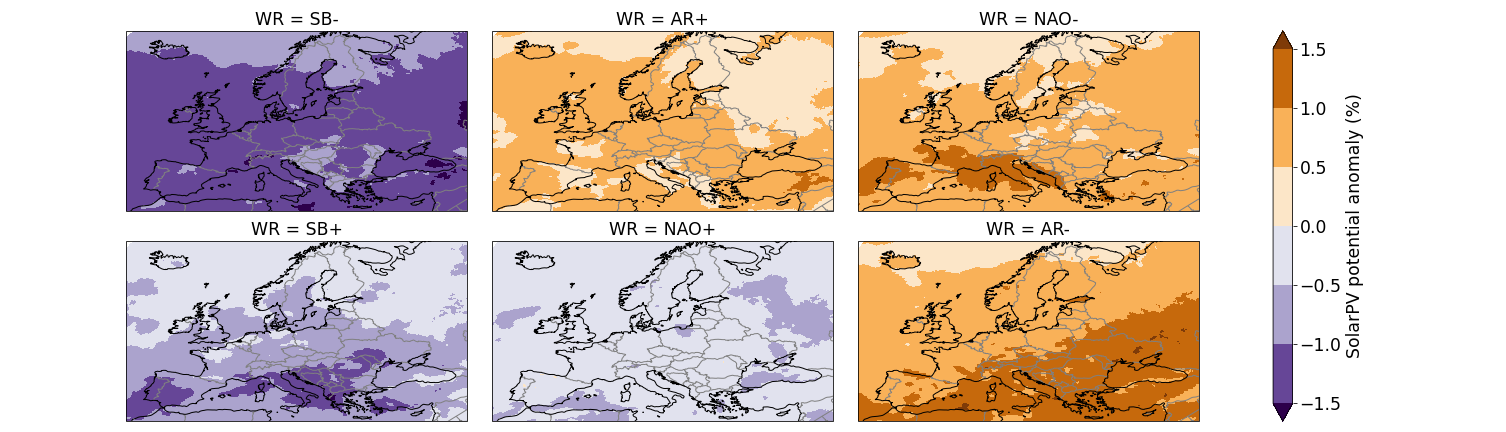}
    \caption{The average solar photovoltaic power generation anomaly from December to March (DJFM) in Europe for each weather regime with respect to the DJFM mean over 1982-2010.}
    \label{fig:WRsolarCFDev}
\end{figure}

An overview of anomaly in the wind speed at 100 meter height, solar irradiance and air temperature at 2 meter height over Europe can be found in Figures \ref{fig:WRwindspeedDev}, \ref{fig:WRSSRDDev}, and \ref{fig:T2mDev}, respectively. The anomaly in solar photovoltaic potential is shown in Figure~\ref{fig:WRsolarCFDev}, the anomaly of the wind turbine potential was provided in Figure \ref{fig:WRwindCFDev}.

\end{document}